\documentclass[fleqn,usenatbib,useAMS]{mnras}

\usepackage{graphicx}	 
\usepackage{amsmath}	  
\usepackage{amssymb}	  
\usepackage{multicol}  
\usepackage{bm}		      
\usepackage{pdflscape}	
\usepackage{hyperref}
\usepackage{xcolor}   

\newcommand{\EBV}{\mbox{$E(4405-5495)$}}
\newcommand{\RV}{\mbox{$R_{5495}$}}

\newcommand{\Teff}{\mbox{$T_{\rm eff}$}}
\newcommand{\alphaM}{\mbox{$\alpha_{\rm M}$}}
\renewcommand{\sci}[2]{\mbox{$#1\cdot 10^{#2}$}}
\newcommand{\nD}{\mbox{$n_{\rm D}(A_1,\alpha)$}}
\newcommand{\ND}{\mbox{$N_{\rm D}(A_1)$}}
\newcommand{\NDp}{\mbox{$N^\prime_{\rm D}(\alpha)$}}
\newcommand{\nM}{\mbox{$n_{\rm M}(A_1,\alpha)$}}
\newcommand{\NM}{\mbox{$N_{\rm M}(A_1)$}}
\newcommand{\NMp}{\mbox{$N^\prime_{\rm M}(\alpha)$}}
\newcommand{\Ns}{\mbox{$N_{\rm s}$}}
\newcommand{\Nr}{\mbox{$N_{\rm r}$}}
\newcommand{\fex}{\mbox{$f_{\rm ex}$}}
\newcommand{\mc}[1]{\multicolumn{2}{c}{#1}}

\usepackage[T1]{fontenc}
\usepackage{ae,aecompl}

\usepackage{newtxtext,newtxmath}

\title[Galactic extinction laws: I. A 2MASS $JHK$ analysis]{Galactic extinction laws: \linebreak
                                                            I. A global NIR analysis with 2MASS photometry}

\author[J. Ma{\'\i}z Apell\'aniz et al.]{J. Ma{\'\i}z Apell\'aniz,$^{1}$\thanks{E-mail: \href{mailto:jmaiz@cab.inta-csic.es}{jmaiz@cab.inta-csic.es}}
M. Pantaleoni Gonz\'alez,$^{1,2}$
R. H. Barb\'a,$^{3}$
P. Garc\'{\i}a-Lario,$^{4}$ and
\newauthor{F. Nogueras-Lara$^{5}$}
\\
$^{1}$Centro de Astrobiolog\'{\i}a. CSIC-INTA. Campus ESAC. Camino bajo del castillo s/n. E-28\,692 Villanueva de la Ca\~nada. Madrid. Spain.\\
$^{2}$Departamento de Astrof{\'\i}sica y F{\'\i}sica de la Atm\'osfera. Universidad Complutense de Madrid. E-28\,040 Madrid. Spain.\\
$^{3}$Departamento de Astronom\'{\i}a. Universidad de La Serena. Av. Cisternas 1200 Norte. La Serena. Chile.\\
$^{4}$European Space Astronomy Centre (ESA/ESAC). Camino bajo del castillo s/n. E-28\,692 Villanueva de la Ca\~nada, Madrid, Spain.\\
$^{5}$Max Planck Institute for Astronomy. K\"onigstuhl 17. D-69\,117 Heidelberg, Germany.
}

\date{Last updated 2020 June 15; in original form 2020 April 7}

\pubyear{2020}

\begin{document}
\label{firstpage}
\pagerange{\pageref{firstpage}--\pageref{lastpage}}
\maketitle

\begin{abstract}
We have started an ambitious program to determine if the full diversity of extinction laws is real or if some of it is due to calibration or methodological issues. 
Here we start by analyzing the information on NIR extinction in a 2MASS stellar sample with good quality photometry and very red colours.
We calculate the extinction at 1~\micron, $A_1$, and the power-law exponent, $\alpha$ ($A_\lambda = A_1 \lambda^{-\alpha}$), for the 2MASS stars located in the 
extinction trajectory in the $H-K$ vs. $J-H$ plane expected for red giants with $A_1 > 5$~mag. We test the validity of the assumption about the nature of those 
stars, whether a single or multiple values of $\alpha$ are needed, and the spatial variations of the results.
Most ($\sim$83\%) of those stars can indeed be explained by high-extinction red giants and the rest is composed of extinguished AGB stars (mostly O-rich),
{blended sources,} 
and smaller numbers of other objects, a contaminant fraction that can be reduced with the help of {\it Gaia}~DR2~data. Galactic red giants 
experience a NIR extinction with $\alpha\sim 2.27$ and an uncertainty of a few hundredths of a magnitude. There is no significant 
spread in $\alpha$ even though our sample is widely distributed and has a broad range of extinctions. Differences with previous results are ascribed to
the treatment of non-linear photometric effects and/or the contaminant correction. 
Future research should concentrate in finding the correct functional form for the NIR extinction law. 
In the appendix we detail the treatment of non-linear photometric effects in the 2MASS bands.
\end{abstract}




\section{Introduction}

$\,\!$\indent The literature on extinction is long and varied and stretches back to a century ago. In modern times, most studies can be divided into two types.
The first one centres on the low-extinction regime ($A_V \lesssim 3$, but see below on the use of band-integrated quantities to characterize extinction) to
study the whole UV/optical/IR ranges simultaneously and derive an overall extinction law. The second one centres on the high-extinction regime 
($A_V \gtrsim 8$), dominant in the Galactic plane, and the IR range, as such stars are weak in the optical and invisible or nearly so in the UV. The first type has the
advantage of the overall approach of the problem but the inconveniences of small samples constrained to the solar neighborhood in the Galactic plane, problematic access 
to the UV, and difficulty of measuring small extinction effects in the IR. The second type has the advantages of large samples and easy to measure effects in the IR but 
the inconvenience of producing results that cannot be easily extrapolated to the optical and UV regimes. Both also suffer from two problems: (a) until recently, most 
studies of one type were disconnected from studies of the other type and (b) there is an over-reliance on using photometry (as opposed to spectrophotometry) for these 
studies, leading to a lack of information on the detailed wavelength-dependent extinction law and possibly to the introduction of systematic biases from an incorrect 
treatment of bandwidth effects (see appendix and references there). The over-reliance on photometry is especially dangerous when the sensitivity curves and/or zero points
are poorly characterized.

Here we start a long-term program to characterize the extinction laws in the Galaxy by addressing some of the issues in the paragraph above. First,
bridging the gap between the two types of papers above by including objects in the intermediate-extinction regime ($3 \lesssim A_V \lesssim 8$) where IR effects
are noticeable and good-quality information in the optical (and even UV) can be obtained. Second, by combining photometry and spectrophotometry to get the best
of both worlds: with large-scale photometric surveys such as {\it Gaia}, Tycho-2, 2MASS, or WISE we get large samples with well-characterized sensitivity curves
and zero-points and with spectrophotometry we get the detailed wavelength behavior. And third, by properly dealing with the non-linear effects of extinction.
We do not have to start from scratch, as we have previously developed tools for this purpose such as CHORIZOS \citep{Maiz04c}, studied the sensitivity curves
and zero points of different photometric systems \citep{Maiz05b,Maiz06a,Maiz07a,Maiz17a,MaizPant18,MaizWeil18}, derived a preliminary family of extinction laws
\citep{Maizetal14a}, analyzed the extinction of O stars in the solar neighborhood \citep{MaizBarb18}, created a spectral energy distribution (SED) model grid
to compare with real data \citep{Maiz13a}, and dealt with the associated non-linear photometric effects (references above plus \citealt{Maiz13b} and the 
appendix here). 

\begin{figure}
\centerline{\includegraphics[width=\linewidth]{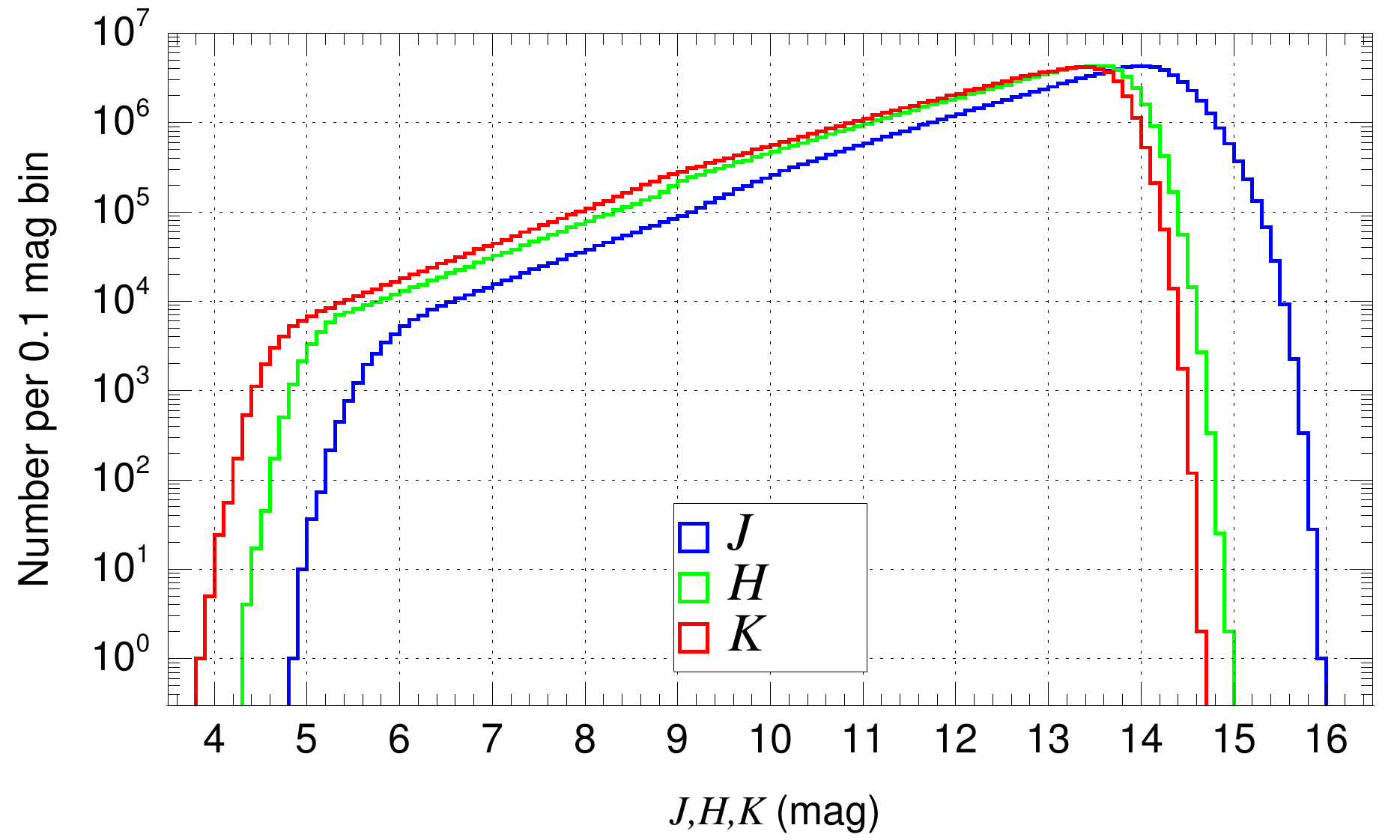}}
\caption{$J$, $H$, and $K$ histograms for the stars in our sample.}
\label{JHK_histo}
\end{figure}

In this paper we analyze the existing 2MASS photometric data to calculate the near infrared (NIR) extinction law assuming its functional form is a power law. In 
future papers we will extend the work to other wavelength ranges and start using spectrophotometry to analyze the detailed wavelength dependence. In the next section 
we describe our methods and data, we then present our results, and we conclude with a comparison with previous works and a description of our future plans. An 
appendix details the non-linear photometric effects that extinction has on 2MASS data.

\begin{figure*}
\centerline{\includegraphics[width=\linewidth]{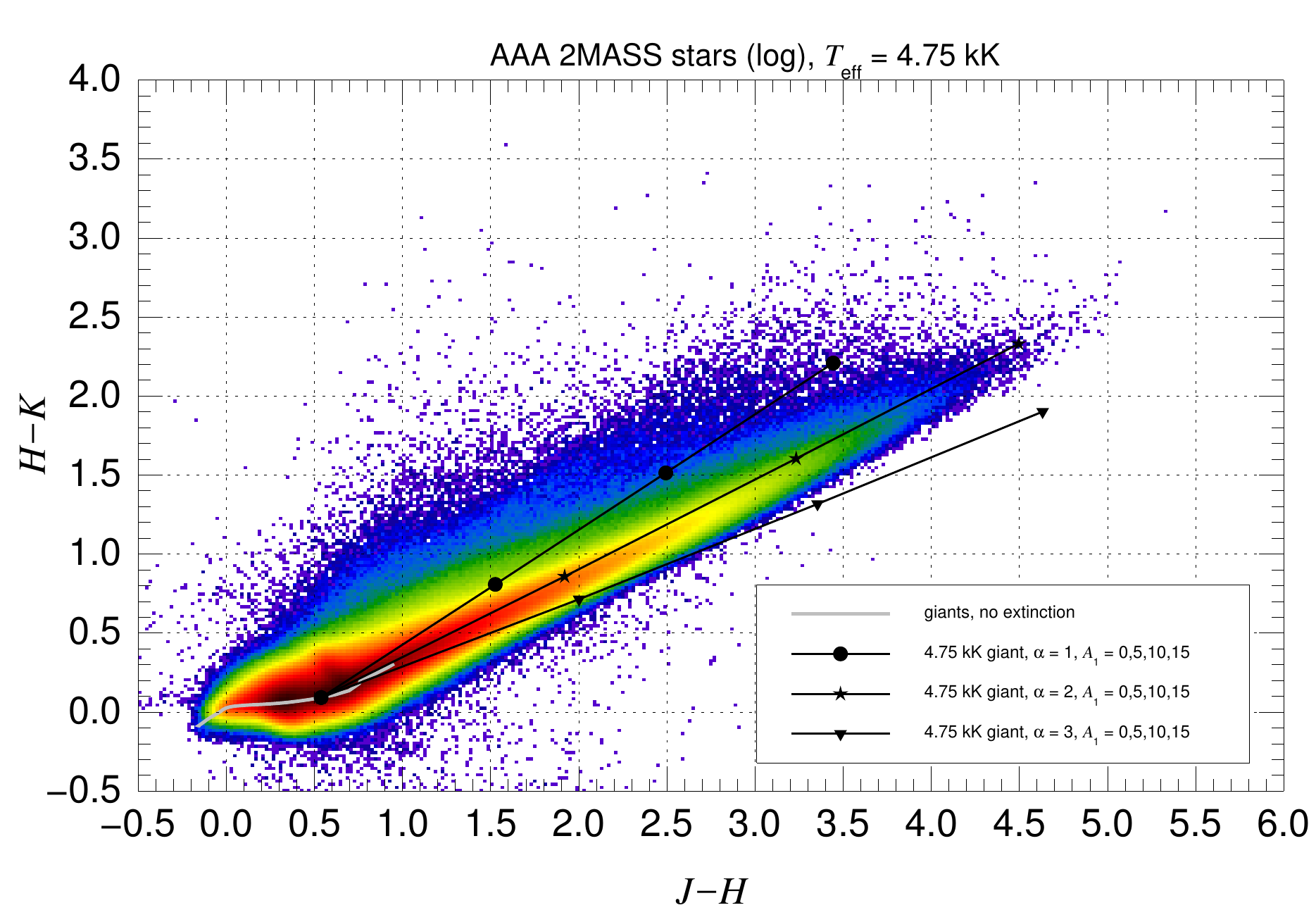}}
\caption{Colour-colour logarithmic-scale density diagram for our full sample using 0.02~mag~$\times$~0.02~mag bins. The gray line is the zero-extinction stellar 
         locus for solar-metallicity giants. The three black lines with symbols show the extinction trajectories for $\alpha$ of 1, 2, and 3 and for the $A_1$ 
         range between 0~and~15~mag for a 4.75~kK giant.}
\label{2MASS_cc_05}
\end{figure*}

\section{Methods and data}

\subsection{Sample}

$\,\!$\indent We obtain our sample from the 2MASS Point Source Catalog \citep{Skruetal06} by selecting all stars with photometric quality flag AAA and with 
uncertainties $\sigma_J\, \le 0.05$, $\sigma_H\, \le 0.05$, and $\sigma_K\, \le 0.05$. The selection cuts are performed in order to retain only those stars
whose photometric precision is good enough to determine their extinction under the conditions described below. The restrictions reduce the number of sources by
a factor of roughly five from the 471 million in the 2MASS PSC to 93 million. A comparison between Fig.~14 in \citet{Skruetal06} (for the region near the north 
Galactic pole) with Fig.~\ref{JHK_histo} here (for the whole sky) reveals the differences in magnitude between the two samples. On the faint end, we reach only 
2.0-2.5~mag brighter than the full sample, as the eliminated stars there do not have enough S/N. On the bright end, our cutoff is at 4-5 mag as brighter sources
are saturated even in the 51~ms exposures (Fig.~17 in \citealt{Skruetal06}) but retained in the 2MASS Point Source Catalog with large uncertainties. 
An $H-K$ vs. $J-H$ density diagram of our sample is shown in Fig.~\ref{2MASS_cc_05}. Our full 93~million sample is later divided in different regions according to 
Galactic coordinates (see below).

\subsection{Calibration and models}

$\,\!$\indent To compare the observed data with synthetic photometry we use the 2MASS bandpasses of \citet{Skruetal06} and the zero points of 
\citet{MaizPant18}. For the stellar models we use the Milky Way metallicity \Teff-luminosity class grid of \citet{Maiz13a}. Details about the specific
SED models used are given in the appendix. In Fig.~\ref{2MASS_cc_05} a grey line is used to place the locus for solar metallicity giants from O stars (left) to 
M stars (right). For other luminosity classes (not plotted), the locus is very similar for hot stars but has significant differences for red ones.
The synthetic photometry is calculated with the package included in CHORIZOS \citep{Maiz04c}, which is written in the IDL 
language\footnote{\url{https://www.harrisgeospatial.com/docs/using_idl_home.html} .}.

\subsection{Extinction treatment}

$\,\!$\indent We assume that the extinction laws in the NIR can be described by a power-law family of the form:

\begin{equation}
A_\lambda = A_1 \lambda^{-\alpha},
\end{equation}

\noindent where $\lambda$ is the wavelength (in microns), $A_1$ is the extinction at 1~micron (in mag), $\alpha$ is a dimensionless exponent that acts as the 
parameter that defines each specific extinction law, and $A_\lambda$ is the extinction at the given wavelength. We employ this form because it is commonly used and
because when one employs a colour-colour diagram (two measured quantities) to determine extinction, as we will do here with 2MASS $H-K$ vs. $J-H$ here, one can 
only fit two parameters (amount $A_1$ and type $\alpha$ of extinction here). More complicated functional forms appear in the literature,
as we will see later, but with 2MASS data alone we only have two data points (colours) so no more parameters can be determined. 
To be strict, we are not ``fitting'' in the sense of using a $\chi^2$ or similar procedure to derive model parameters from a data set with a higher dimension
but, as the number of points corresponds to the number of parameters, we are solving a set of equations or, in geometrical terms, we are doing a one-to-one
mapping between a location in the $H-K$ vs. $J-H$ plane and a location in the $\alpha$ vs. $A_1$ plane (with underlying assumptions about the input SED, see
below).

The lower left part of Fig.~\ref{2MASS_cc_05}, where most of the 2MASS point sources are, is populated mostly by low-extinction main-sequence stars, which 
will be ignored in this paper as when they are extinguished they become to dim to be present in 2MASS with good photometric quality. Our interest is in the sequence 
that moves diagonally from the lower left to the upper right of the $H-K$ vs. $J-H$ diagram and that is mostly an extinction line for red giants \citep{Comeetal02}. 
Among those, the majority are red clump (RC) stars \citep{Gira16} with a minority of more luminous objects
in the red giant branch (RGB) \citep{StraLaug08c}. There is also an additional population above the main extinction sequence for red giants with two main components: 
(a) Asymptotic giant branch (AGB) stars, which have extreme NIR colours and can also be observed through long sightlines across the Galaxy and (b) OB stars, which are 
intrinsically bluer and more scarce than red giants, but also luminous and concentrated (even more so than red giants) in the Galactic plane. Additional populations 
such as red supergiants, Wolf-Rayet stars, or young stellar objects are more scarce.

An important point here is that we can analyze the extinction sequence in the $H-K$ vs. $J-H$ colour-colour diagram by assuming that it is composed of
extinguished RC stars with a small fraction of contaminants. Therefore, we will start with the hypothesis that all objects there are a single-temperature
population, see how that hypothesis is consistent with the data, and introduce modifications to determine the robustness of our results and the proportion of
contaminants. Galactic red clump stars have a range of \Teff\ between 4.5~and~5.0~kK \citep{Bovyetal14}, which in the MW metallicity giant models of \citet{Maiz13a} 
correspond to a range of $(J-K)_0$ between 0.55 and 0.73. We therefore choose as our base model the intermediate 4.75~kK one, which has 
$(J-K)_0 = 0.63$, a value quite similar to the empirical one of 0.68 by \citet{Gonzetal12}.

\begin{figure}
\centerline{\includegraphics[width=\linewidth]{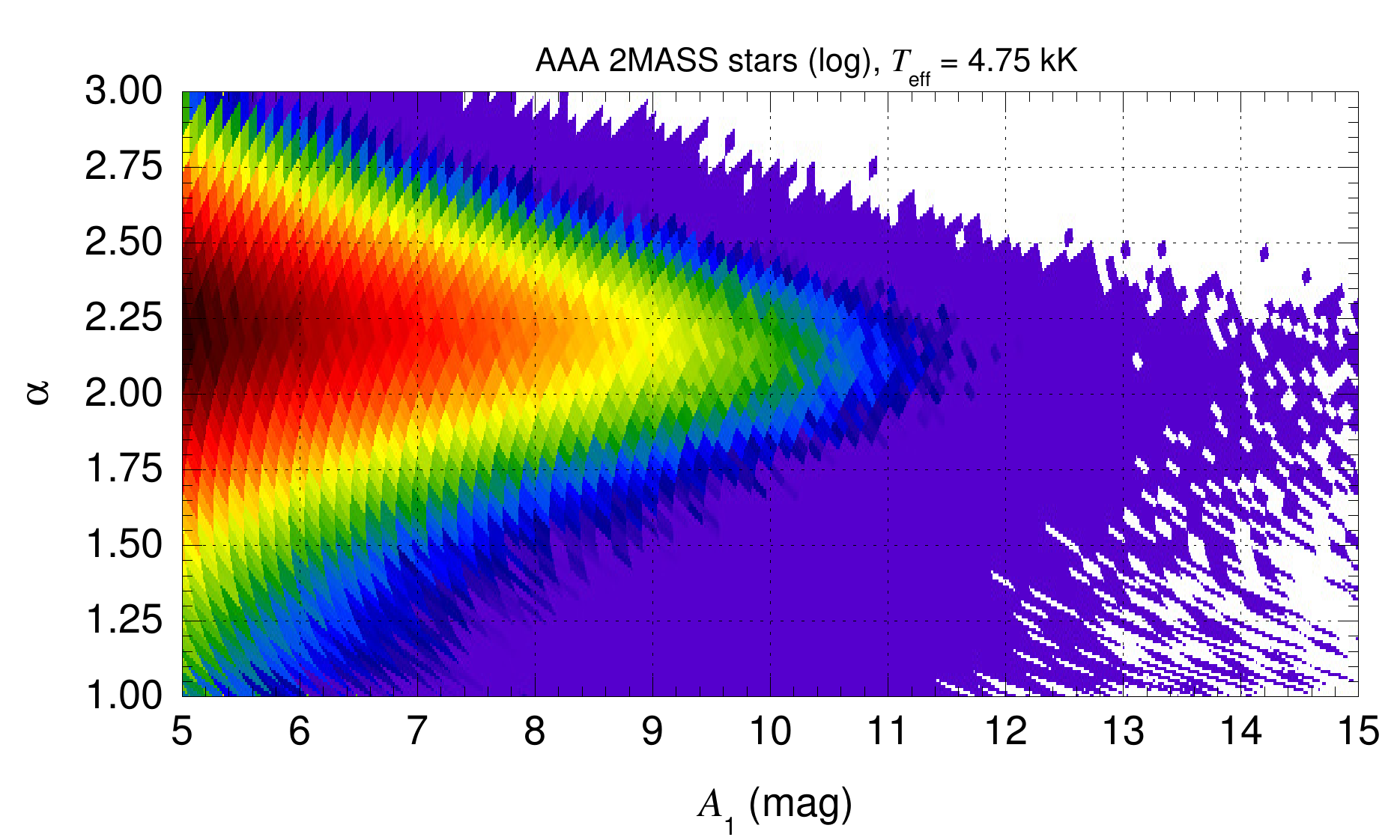}}
\vspace{-2mm}
\centerline{\includegraphics[width=\linewidth]{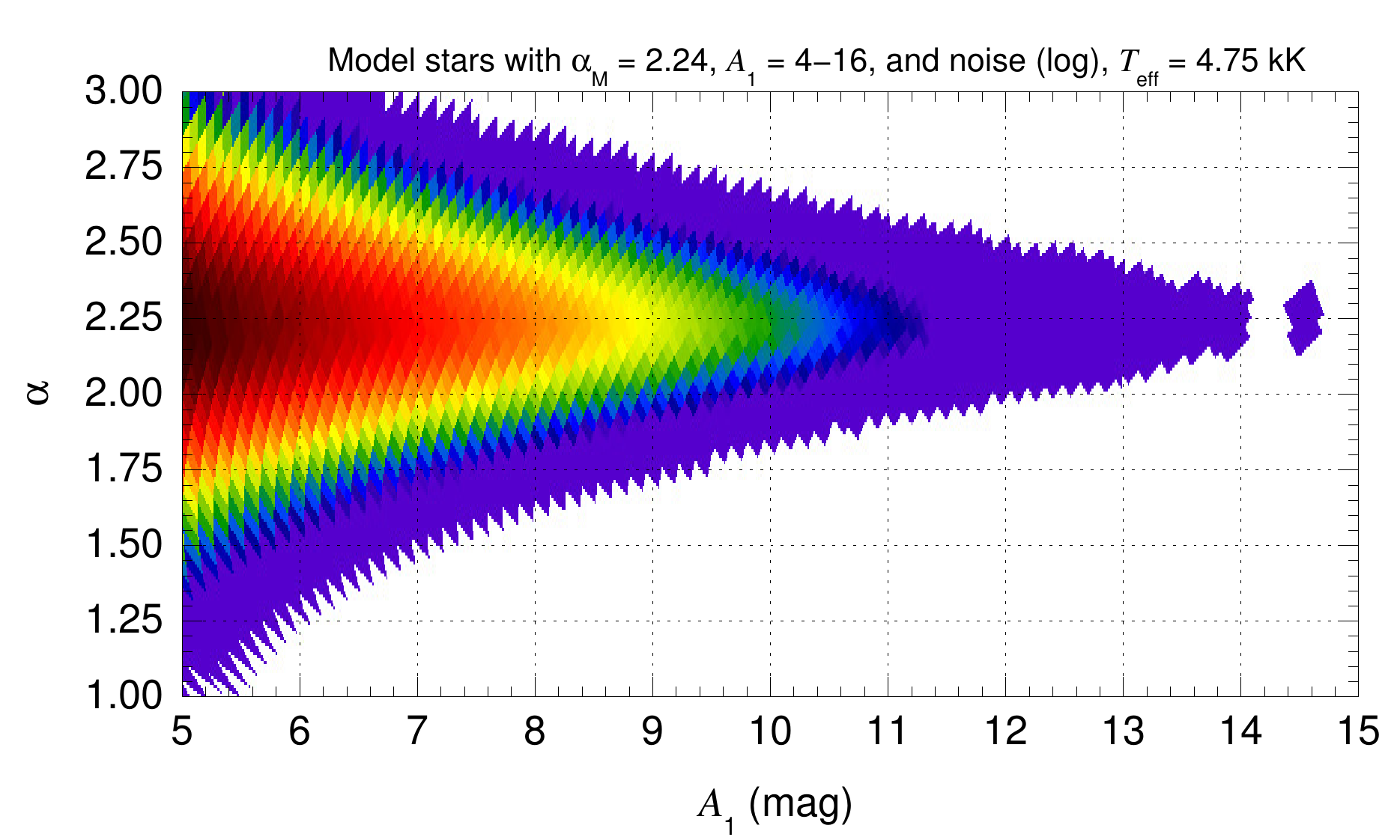}}
\vspace{-2mm}
\centerline{\includegraphics[width=\linewidth]{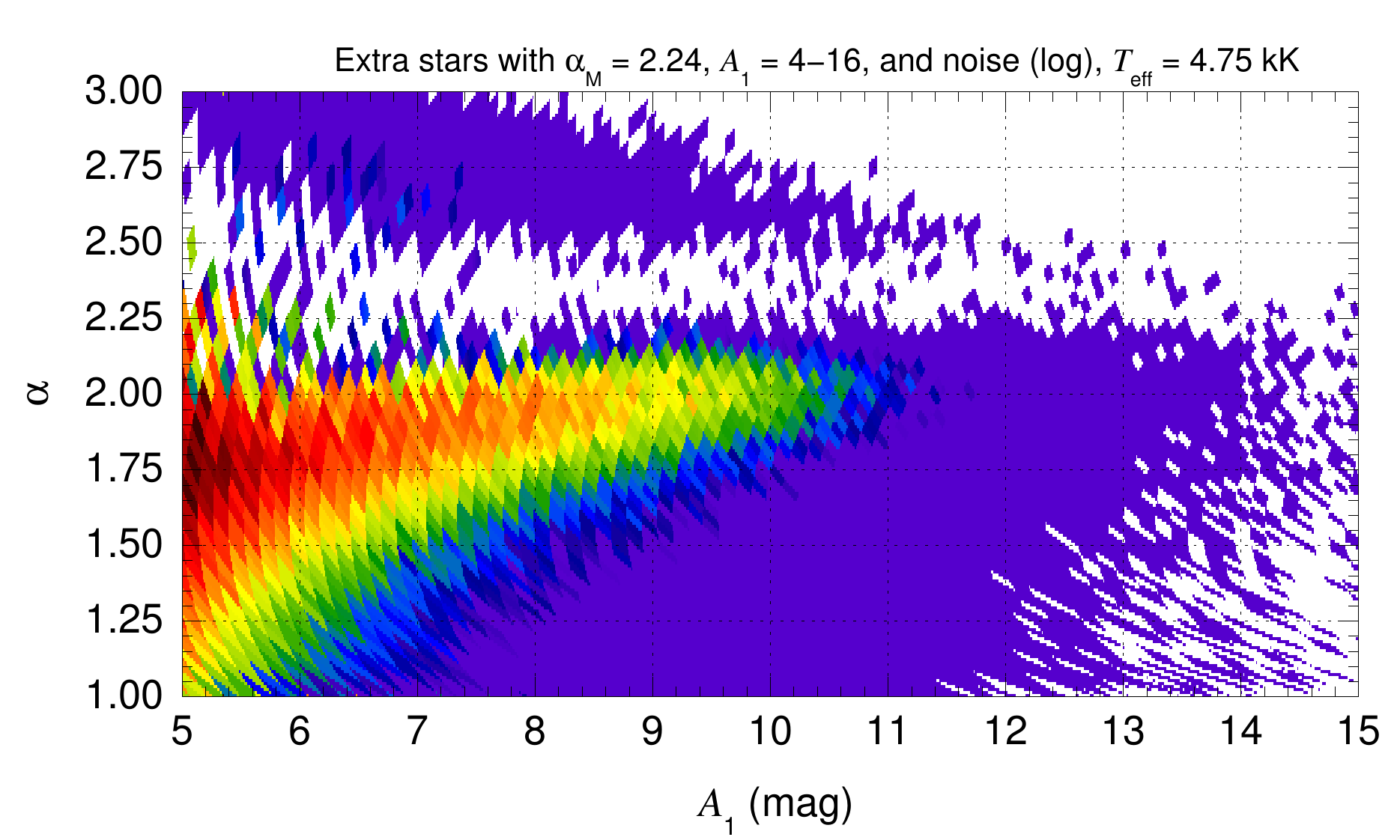}}
\vspace{-2mm}
\centerline{\includegraphics[width=\linewidth]{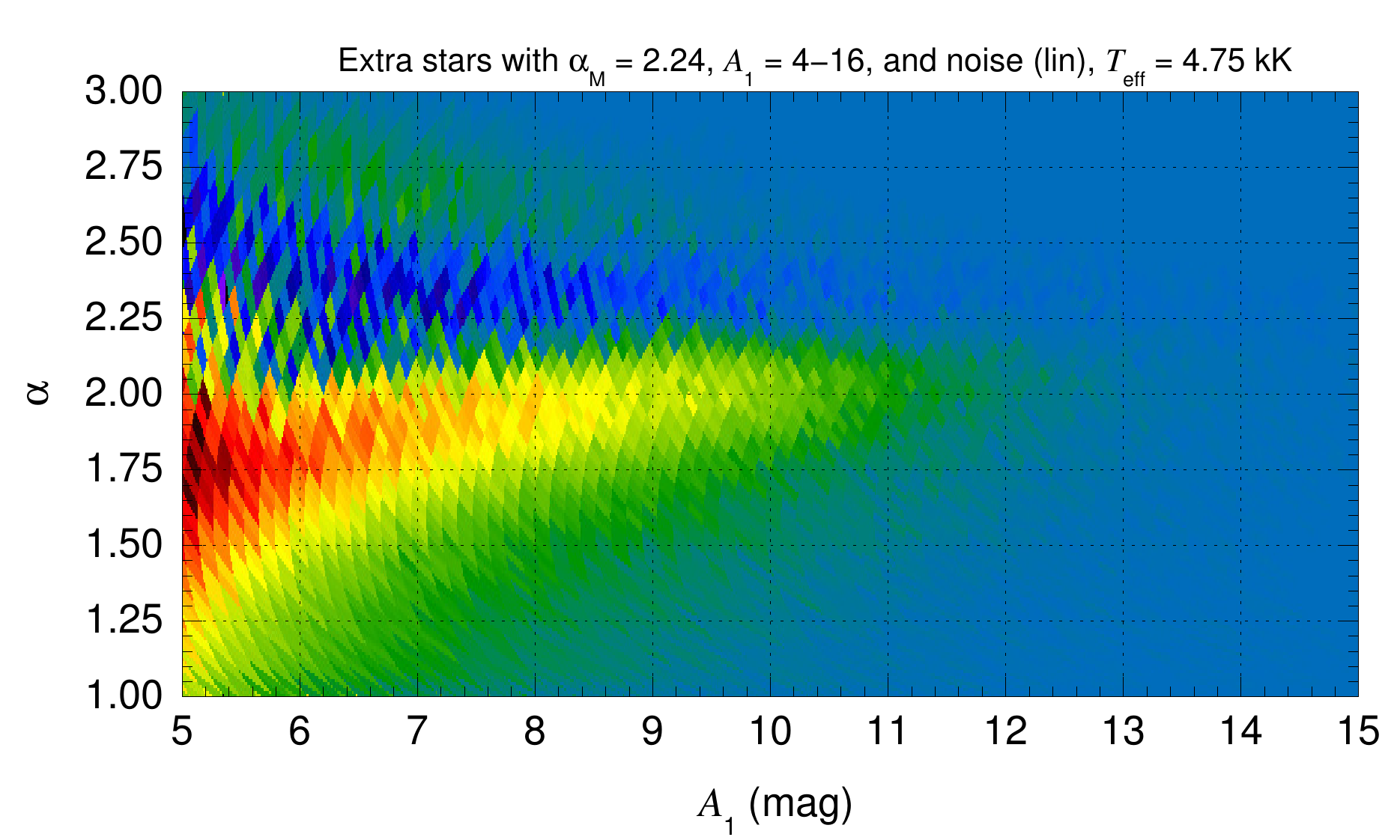}}
\caption{$A_1$-$\alpha$ density diagrams for the whole sample assuming that all stars are 4.75~kK giants with solar metallicity. The top panel shows the data 
         (\nD, logarithmic scale), the next panel the model assuming \alphaM~=~2.24 plus noise (\nM, logarithmic scale), and the bottom panels the 
         difference between the two (logarithmic scale with positive values and linear scale with positive and negative values).}
\label{2MASS_par_05_4750}
\end{figure}

A fundamental point in this paper is that {\bf the model SED is extinguished first and the synthetic photometry is computed on the extinguished SED to account
for non-linear extinction effects}, as described in detail in the appendix. For a given input SED we do that for a range of $A_1$ values between 0~and~16~mag 
and a range of $\alpha$ values between 1~and~3, which provides us a with a mapping between a given value in the region of the $H-K$ vs. $J-H$ plane in 
Fig.~\ref{2MASS_cc_05} enclosed by the extinction trajectories plotted there and the corresponding value in the $A_1$ vs. $\alpha$ plane. This mapping yields 
the 2-D data density distribution \nD\ and the correspondence can be seen by comparing Fig.~\ref{2MASS_cc_05}
with the top panel of Fig.~\ref{2MASS_par_05_4750}, with two clarifications: (a) In Fig.~\ref{2MASS_par_05_4750} (and
subsequent ones) we restrict the range of $A_1$ to 5-15 mag, as lower values are more heavily contaminated from other types of stars and for higher values there
are very few stars in our sample. (b) The mapping from one plane to one another introduces a distortion in the original square cells; we have chosen to keep
the distorted shapes in Fig.~\ref{2MASS_par_05_4750} to guide the eye with respect to cell size (0.02~mag~$\times$~0.02~mag in $H-K$ vs. $J-H$) and orientation.

Next, we simulate a single-temperature model that describes the observed distribution in the top panel of Fig.~\ref{2MASS_par_05_4750}.
We do that in the following steps.

\begin{itemize}
 \item We produce a first estimate of the number of stars per amount of extinction $N_{\rm M,0}(A_1)$ by integrating over the vertical direction of the top panel 
       of Fig.~\ref{2MASS_par_05_4750} to obtain \ND, extending the limits to 4-16~mag to include the behavior close to the edges.
 \item We calculate the average uncertainties $\sigma_J$, $\sigma_H$, and $\sigma_K$ in the $H-K$ vs. $J-H$ plane and map them into the $\alpha$ vs. $A_1$ 
       plane to calculate $\sigma_J(A_1)$, $\sigma_H(A_1)$, and $\sigma_K(A_1)$. This is done for accuracy purposes but is should be noted that those values are
       nearly constant in the range of interest in Fig.~\ref{2MASS_cc_05}. More specifically, $\sigma_J = 32.4\pm 5.6$~mmag, $\sigma_H = 30.4\pm 5.5$~mmag, 
       and $\sigma_K = 27.8\pm 6.6$~mmag there.
 \item Once we have $N_{\rm M,0}(A_1)$ and $\sigma_J(A_1)+\sigma_H(A_1)+\sigma_K(A_1)$, we estimate an initial model $\alpha_{\rm M,0}$ (i.e. 2.0) and apply the 
       dispersion associated with the uncertainties to produce a 2-D distribution $n_{\rm M,0}(A_1,\alpha)$. Three important points have to be made here. The 
       spread in $\alpha$ in the 2-D distribution is not real but an artifact of the magnitude uncertainties; in reality we are assuming a single 
       $\alpha_{\rm M,0}$ value. Second, the uncertainties in $J$, $H$, and $K$ are independent but those of $J-H$ and $H-K$ are anticorrelated, see the ellipses 
       in Fig.~2 of \citet{MaizPant18}\footnote{But note that there is a typo in the caption there, as those ellipses correspond to uncertainties of 15~mmag, not 
       1.5~mmag.}.  Third, objects near the edges (mostly the left one) are moved in and out of the plot by the application of the procedures (edge diffusion). 
       That is why we extend the $A_1$ limits to 4-16~mag, even though we only analyze the 5-15~range.
 \item The resulting distribution $n_{\rm M,0}(A_1,\alpha)$ is compared with the observed data by (a) changing the value of $\alpha_{\rm M,0}$, (b) tweaking 
       $N_{\rm M,0}(A_1)$ to account for edge diffusion, and (c) iterating until we are satisfied with the final result: a model NIR slope, \alphaM, 
       an observed model distribution, \nM, and total number of stars as a function of amount of extinction, \NM, and as a function of $\alpha$, \NMp.
\end{itemize}

\begin{figure}
\centerline{\includegraphics[width=\linewidth]{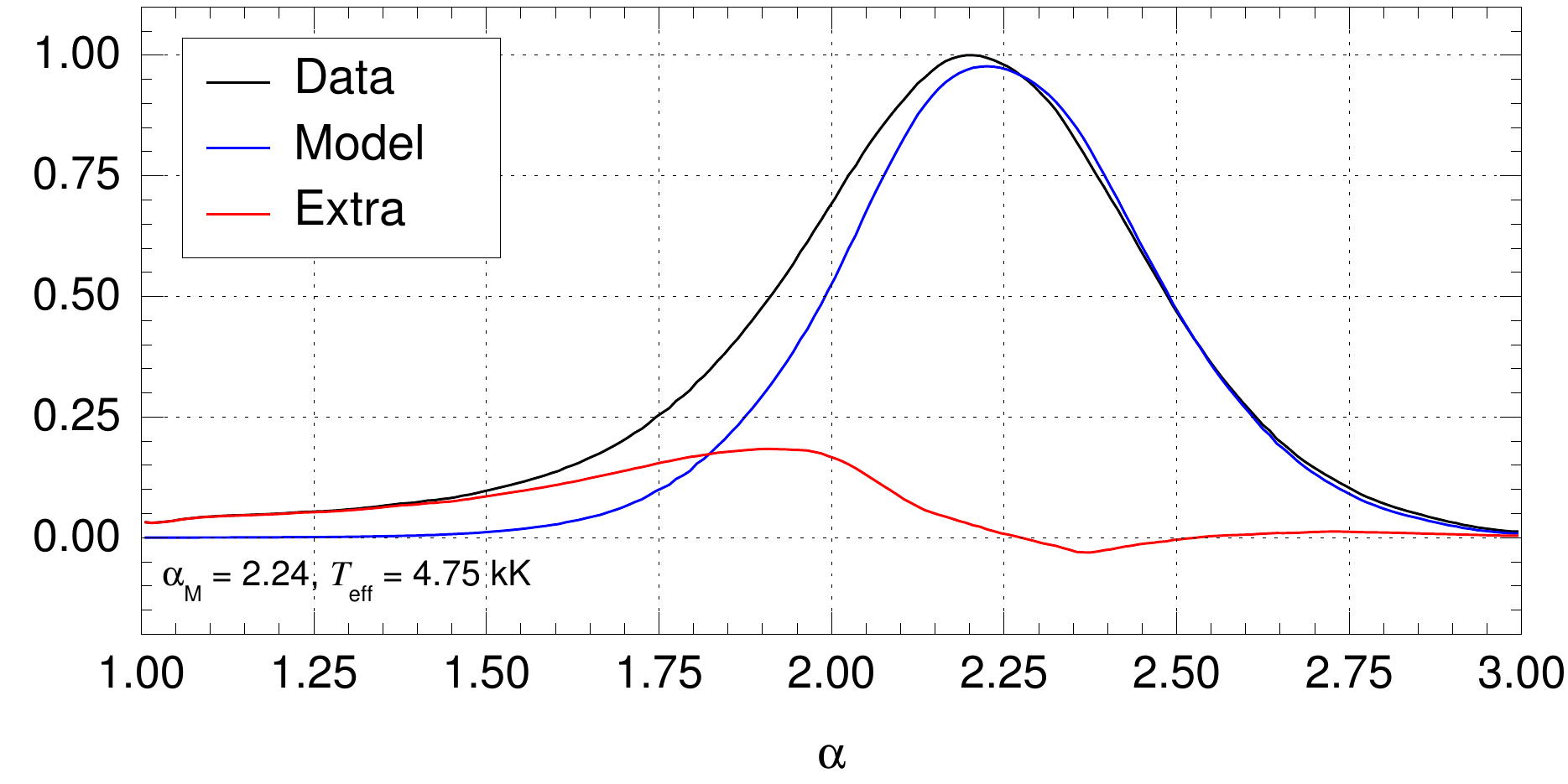}}
\centerline{\includegraphics[width=\linewidth]{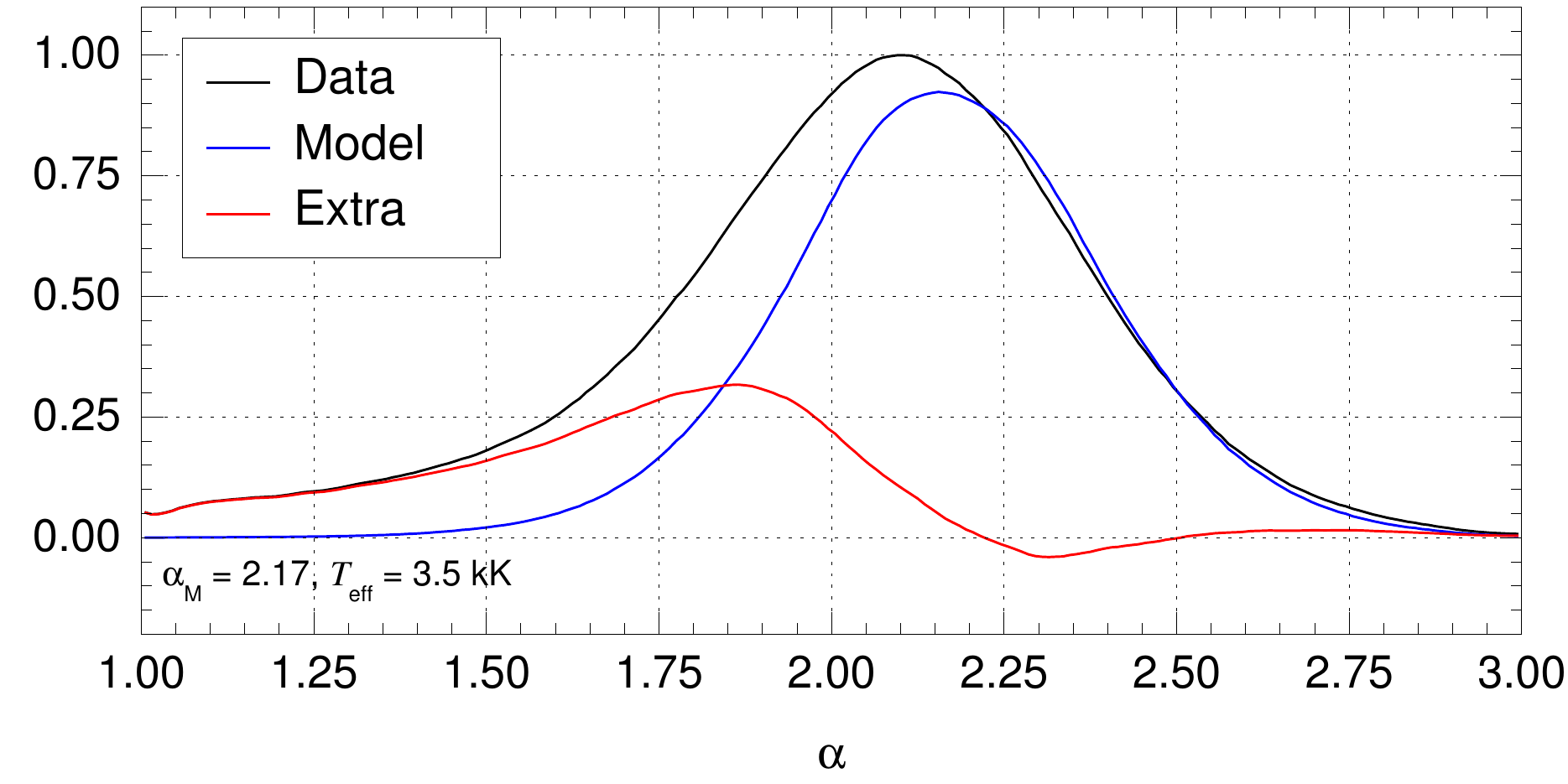}}
\centerline{\includegraphics[width=\linewidth]{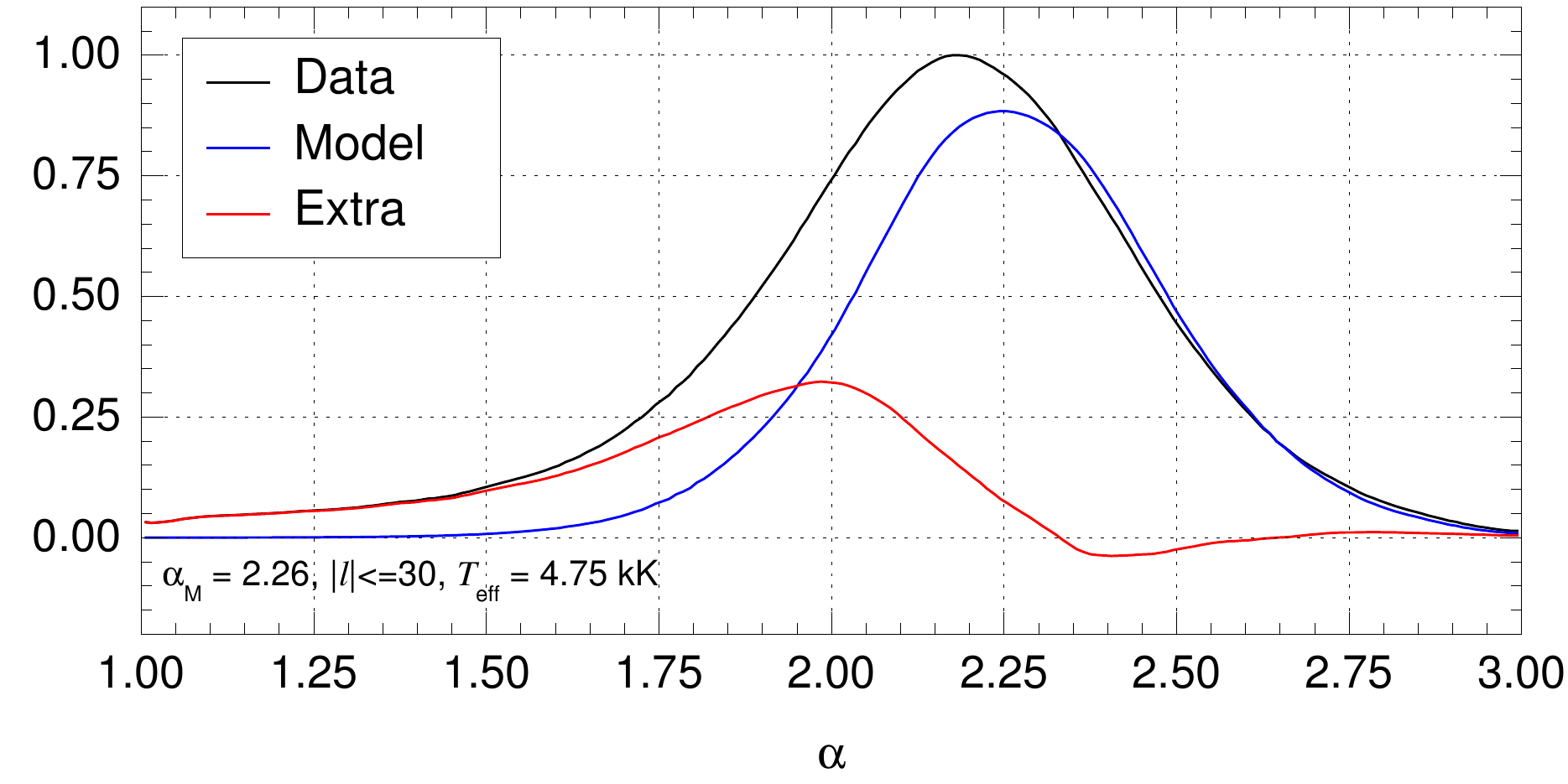}}
\centerline{\includegraphics[width=\linewidth]{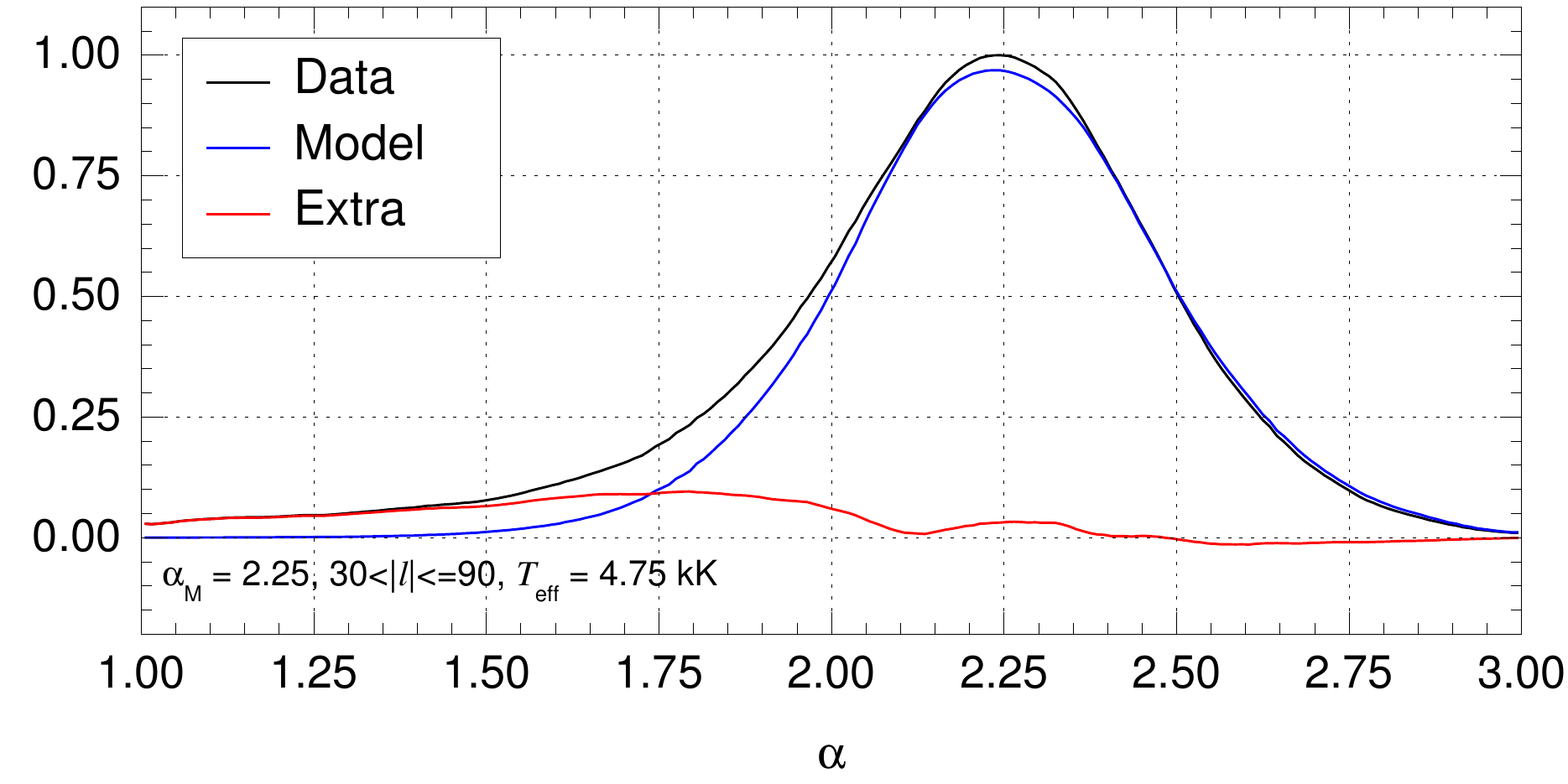}}
\centerline{\includegraphics[width=\linewidth]{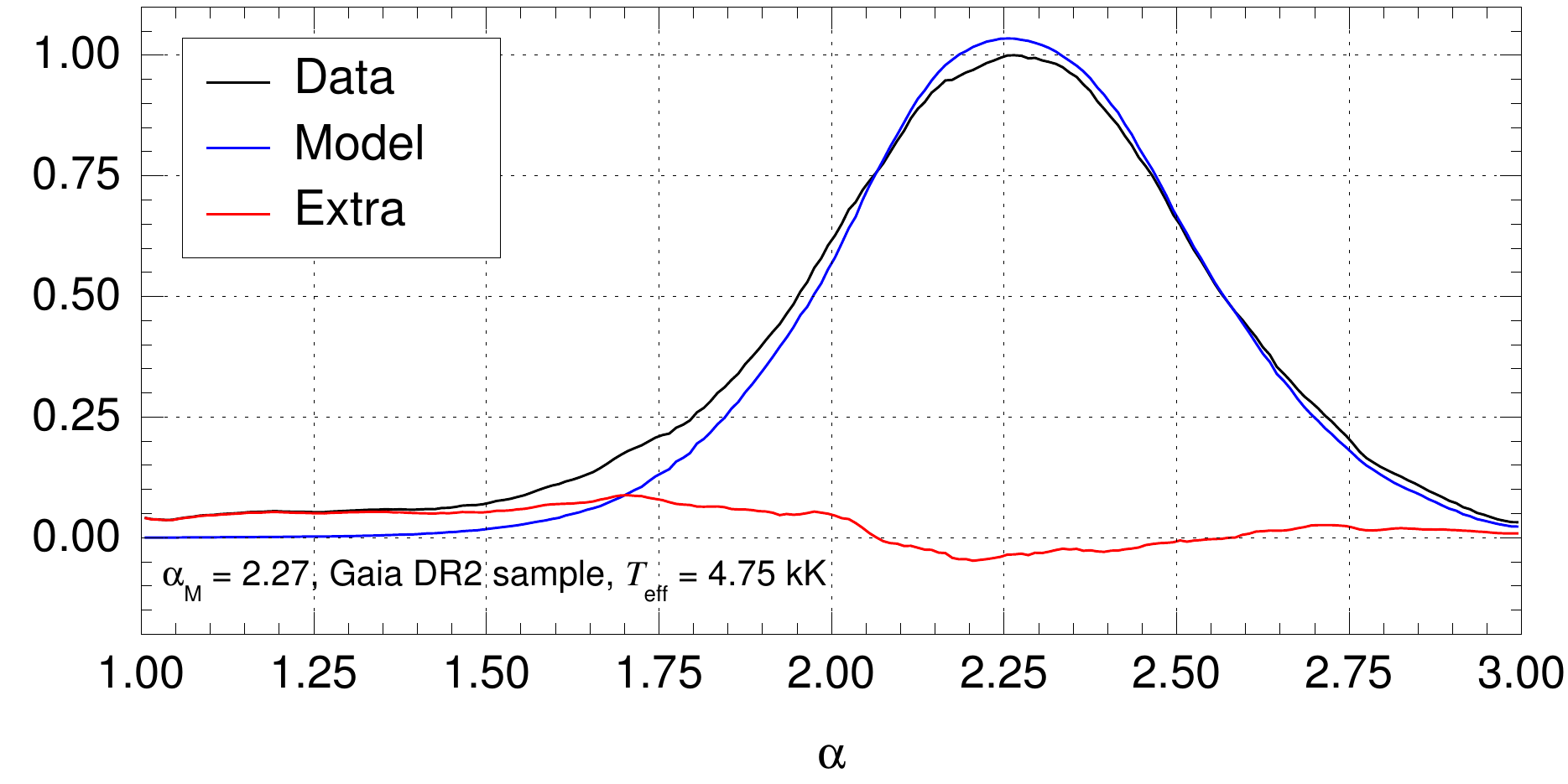}}
\caption{Normalized $\alpha$ density distribution for the data, \NDp; model, \NMp; and their differences for different experiments. The top two panels
         are for the whole sample assuming that stars are 4.75~kK (first) or 3.5~kK (second) giants with solar metallicity. The next two panels
         consider the central (third) and lateral (fourth) Galactic regions, also for 4.75~kK giants. The last panel is the equivalent for the {\it Gaia}~DR2
         selected sample.}
\label{alpha_density}
\end{figure}

As described below, the procedure above is executed first for the whole sample using two different values of \Teff\ and then for different subsamples defined
from regions in the sky and {\it Gaia}~DR2 data. For the whole sample case assuming a \Teff\ of
4.75~kK we also plot in Fig.~\ref{2MASS_par_05_4750} the model \nM\ and the difference between data and model or ``extra stars'' (in both logarithmic and linear 
scales with different ranges). In addition, the plots in Fig.~\ref{alpha_density} show \NMp, its data equivalent \NDp, and the difference between the second and
the first. Results are given in Table~\ref{results}, where \Ns\ is the 
number of stars in the specific sample, \Nr\ is the number in the $A_1$~=~5-15~mag + $\alpha$~=~1-3 range, and \fex\ is the percentage of 
\Ns\ not included in the model (the area under the red curve divided by the area under the black curve in the top panel of Fig.~\ref{alpha_density}).

\begin{table}
\caption{Results for different samples: numbers, assumed \Teff, and measured parameters.}
\label{results}
\begin{tabular}{cccccr}
sample              & \Ns\          & \Nr\          & \Teff & \alphaM & $f_{\rm ex}$ \\
                    &               &               & (kK)  &         & (\%)         \\
\hline
whole               & \sci{9.31}{7} & \sci{1.08}{6} & 4.75  & 2.24    & 17.0         \\
whole               & \sci{9.31}{7} & \sci{4.62}{5} & 3.50  & 2.17    & 27.4         \\
$|l|\, \le 30$      & \sci{3.11}{7} & \sci{7.46}{5} & 4.75  & 2.26    & 27.5         \\
$30 < |l|\, \le 90$ & \sci{3.53}{7} & \sci{3.34}{5} & 4.75  & 2.25    & 11.3         \\
{\it Gaia} DR2      & \sci{1.55}{5} & \sci{2.16}{4} & 4.75  & 2.27    &  7.6         \\
\hline
\end{tabular}
\end{table}

\section{Results}   

\subsection{Whole sample, \Teff\ = 4.75 kK}

$\,\!$\indent We start with an analysis of the whole sample assuming it is composed of \Teff~=~4.75~kK solar-metallicity giants. The first result is that \NDp\ 
is quasi-gaussian 
{(meaning a symmetric single-peaked distribution with kurtosis close to 3) but not truly} 
so, as there is a wing that extends towards lower values of $\alpha$ (or values of $H-K$ larger that what would correspond for a
given $J-H$ along an extinction line for the most common $\alpha$). On the other hand, \NMp\ is quasi-gaussian, as a result of the spread in $J-H$ and $H-K$
following a gaussian (correlated) distribution and the mapping between the colour-colour and the $A_1-\alpha$ plane introducing only weak distortions. The existence
of the wing, the main difference between \NDp\ and \NMp, has at least 
{four} 
possible explanations: [a] a subpopulation of \Teff~=~4.75~kK solar-metallicity 
giants extinguished with a real lower value of $\alpha$ i.e. a spread in the NIR extinction law properties; [b] red giants with a lower \Teff\ but with the same 
or similar $\alpha$; [c] contamination from other types of stars such as AGB stars or extinguished OB stars; 
{and [d] source blending in crowded regions, which is more likely to make sources brighter in $K$ than in $J$ or $H$.} 
In the following subsections we make different assumptions and select subsamples to test the validity of those explanations. 

The value we obtain for \alphaM\ is 2.24 and the wing contains 17\% of the sample. It should be noted that, in principle, it is possible to obtain alternative 
fits by reducing the value of \alphaM. If this is done, several things happen. First, the value can only be reduced by a small amount, down
to a minimum \alphaM\ of 2.18. Second, if that is done, we are left with wings on both sides, as the whole width of \NDp\ (determined by the photometric
uncertainties alone for a single \Teff\ population extinguished with a single $\alpha$) cannot be explained by a single value of \alphaM, i.e. one partial 
solution would be to have a distribution of values of $\alpha$ around 2.20 (which is another way of rephrasing option [a] above). Third, even if we adopt such a 
spread, \NDp\ is clearly asymmetric and the tail extends beyond $\alpha = 1$, which points towards option [c] above as having a non-negligible contribution. 
Therefore, our first result is that a possible explanation for \nD\ is a dominant population of RC stars extinguished with $\alpha$ around 2.24 (with a possible
small spread around that value) but with a secondary population that constitutes a small fraction of the sample and whose nature will be explored in the 
following subsections.

\subsection{Whole sample, \Teff\ = 3.5 kK}

$\,\!$\indent As a second experiment we analyze the possible contribution of cooler, more luminous stars from the RGB, which we know must be present in the 
sample to some extent \citep{StraLaug08c}. We repeat the previous experiment substituting our model by a 3.5~kK giant solar metallicity model from
\citet{Maiz13a}. Results are shown in Fig.~\ref{alpha_density} and Table~\ref{results}.

This experiment yields results that are similar to the previous one but with some significant differences. (a) \Ns\ has been reduced by more than one half because
as RGB stars are intrinsically redder than RC stars, the same $A_1$ corresponds to redder colours and the sample within the range being studied is smaller. (b)
\alphaM\ is slightly smaller (2.17 vs. 2.24) because RGB stars are located very slightly below the extinction trajectory of a RC star with $\alpha = 2.24$ in the 
$H-K$ vs. $J-H$ plane. (c) \fex\ is slightly larger, as the low-$\alpha$ wing contains a larger area.

The interpretation of the above is that RGB stars may be contaminating our sample but the main effect is one of simply underestimating their amount of extinction
when we assume that their \Teff\ is higher than their real value while introducing slight changes of a few hundredths in the value of $\alpha$. They may produce a 
slight spread in the observed distribution of $\alpha$ but nowhere close to explaining the low-$\alpha$ wing, whose origin must be interpreted in a different way.
In summary, tweaking the characteristics of the red giants in the sample only produces small changes in \NDp\ and switching from RC to RGB stars just
makes the wing stronger.

\subsection{Moving around the Galactic plane, \Teff\ = 4.75 kK}

$\,\!$\indent We now compare different regions in the sky. We define the central region of the Galaxy as that with $|l|\, \le\, 30^{\rm o}$ (for 
ease of notation we consider $l$ to be between $-180^{\rm o}$ and $180^{\rm o}$) and the rest of the inner two quadrants (``lateral'') as the region with
$30^{\rm o} < |l|\, \le 90^{\rm o}$, with both regions having a similar \Ns. In the central region the subsample is likely dominated by Galactic bulge stars but 
the disc has a significant contribution while in the lateral region most high-extinction objects are Galactic disc stars. We ignore the two outer quadrants, as 
extinction there is significantly lower in our 2MASS sample. For the central and lateral regions, results are given in Table~\ref{results} and plotted in 
Fig.~\ref{alpha_density}. Note that most of the 2MASS sources in general are concentrated towards the Galactic plane and that is even more so for highly extinguished 
ones. There are very few objects with $|b| > 5$ that have $A_1 > 5$~mag in any of our experiments, so for practical purposes restricting by longitude means studying 
different parts of the Galactic plane and the nearby bulge.

For both regions the value of \alphaM\ is very similar to that of the first experiment, indicating there is no large variation of $\alpha$ with Galactic 
longitude. However, one important difference appears: \fex\ is more than double in the central region than in the lateral one (with the value from the first
experiment in between), indicating that either (a) the central region contains a scatter in $\alpha$ with values in the approximate range 2.10-2.25 or (b) the 
contaminant population is centrally concentrated. We explore the nature of that population next.

\subsection{What is the contaminant population?}


\begin{figure}
\centerline{\includegraphics[width=\linewidth]{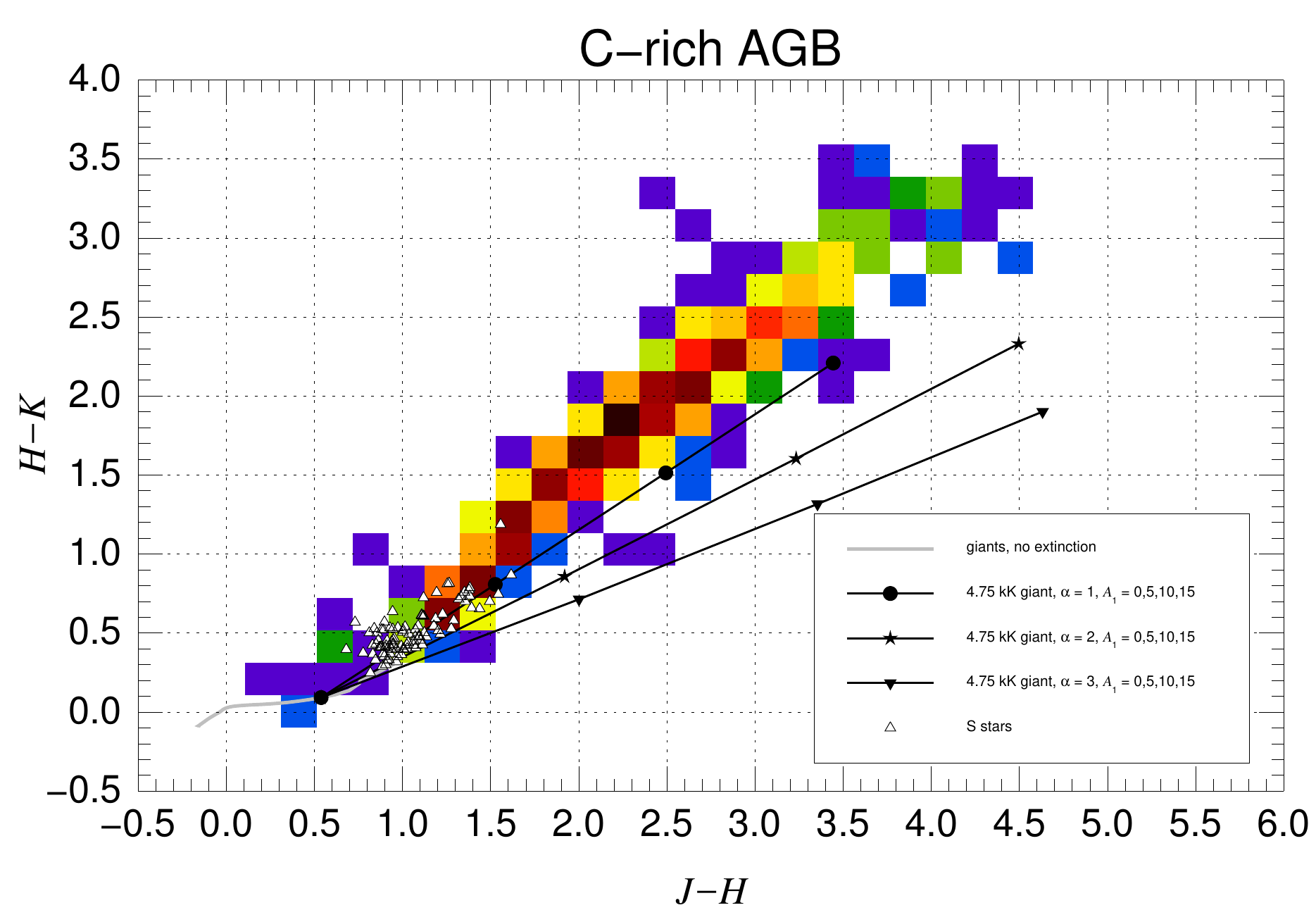}}
\centerline{\includegraphics[width=\linewidth]{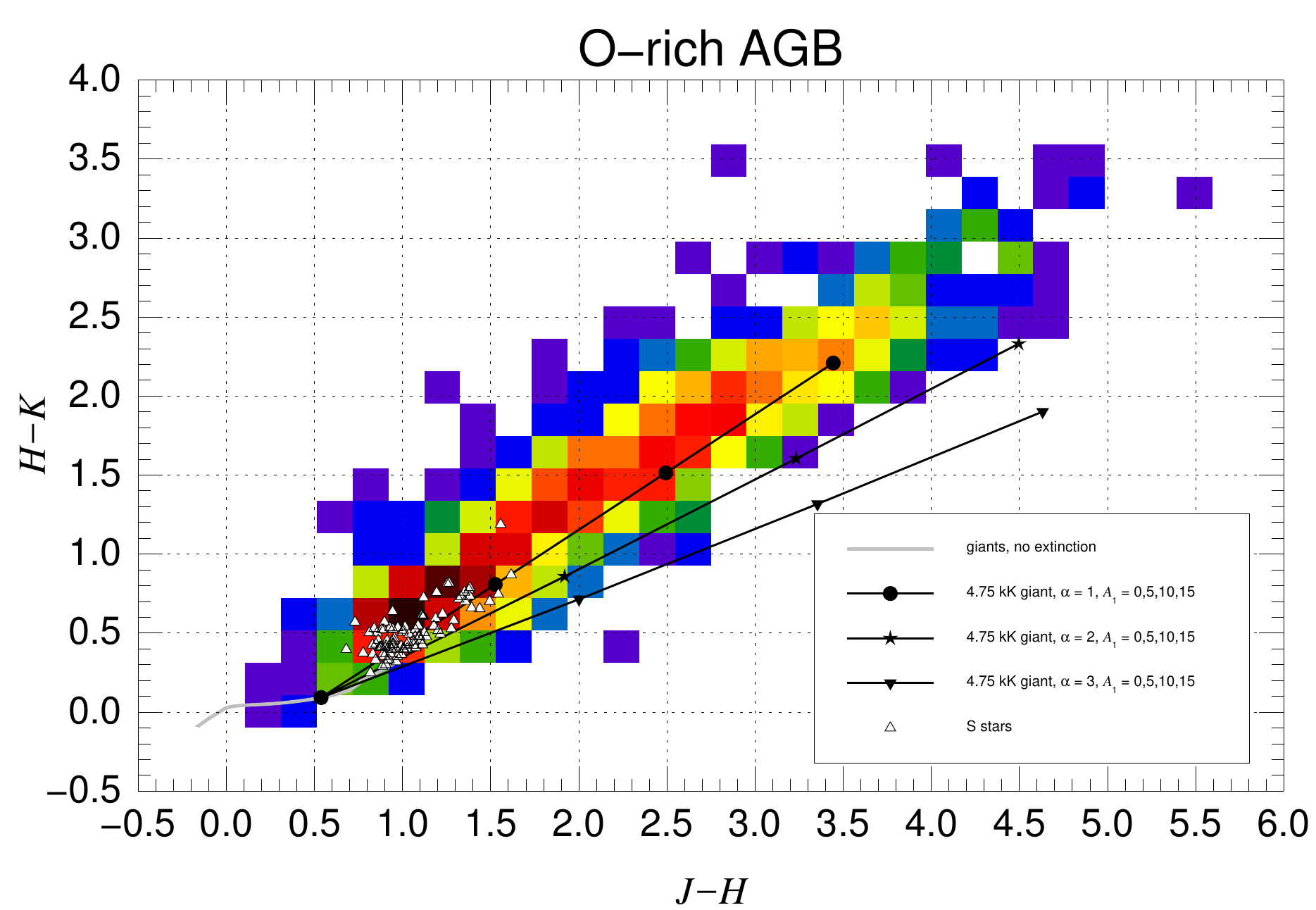}}
\centerline{\includegraphics[width=\linewidth]{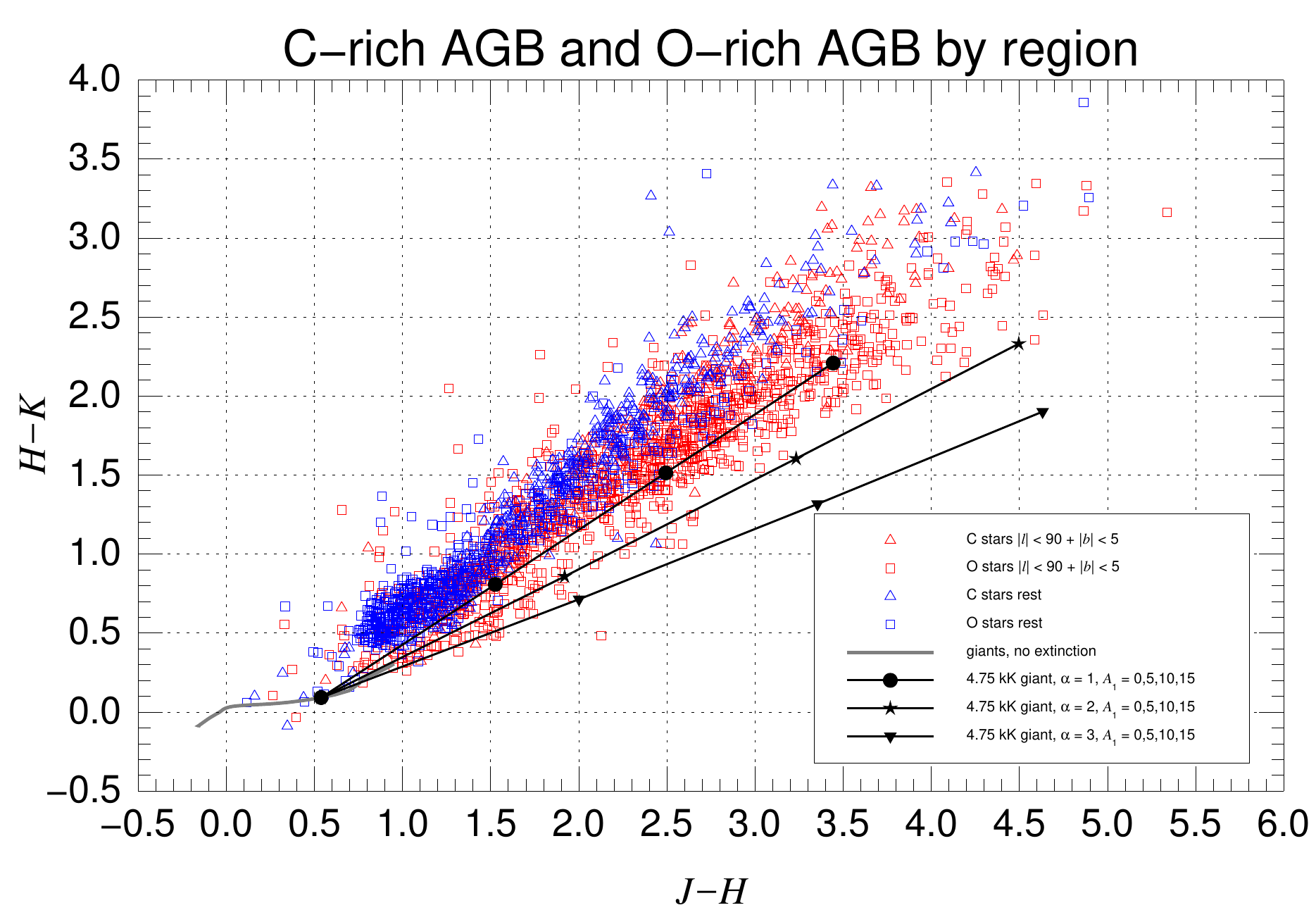}}
\caption{Colour-colour logarithmic-scale density diagram for Galactic C-AGB (top) and O-AGB (middle) stars with S stars added as symbols in both plots.
         The bottom plot shows the individual C-AGB and O-AGB stars divided in two regions: the Galactic plane in the inner two quadrants and the rest of the
         sky. The gray line is the zero-extinction stellar locus for solar-metallicity giants. The three black lines with symbols show the extinction 
         trajectories for $\alpha$ of 1, 2, and 3 and for the $A_1$ range between 0~and~15~mag for a 4.75~kK giant.}
\label{2MASS_cc_AGB}
\end{figure}

\begin{table*}
\caption{Statistics for previously known AGB stars analyzed in this paper.}
\label{results_AGB}
\centerline{
\begin{tabular}{lccr@{$\pm$}lr@{$\pm$}lr@{$\pm$}lr@{$\pm$}lr@{$\pm$}lr@{$\pm$}l}
type      & total & $f_{\rm IGP}$ & \mc{$J$}   & \mc{$J$}   & \mc{$J-H$} & \mc{$J-H$} & \mc{$H-K$} & \mc{$H-K$} \\
          &       & (\%)          & \mc{rest}  & \mc{IGP}   & \mc{rest}  & \mc{IGP}   & \mc{rest}  & \mc{IGP}   \\
\hline
C-AGB     &  621  & 38.3          &  9.76&2.32 & 10.77&2.20 & 2.07&0.70  & 2.53&0.75  & 1.60&0.68  & 1.89&0.67  \\ 
O-AGB     & 1800  & 61.9          &  8.29&2.17 & 10.82&2.49 & 1.43&0.67  & 2.28&0.87  & 0.98&0.53  & 1.49&0.64  \\ 
S stars   &   94  & 29.8          &  6.80&0.87 &  6.84&0.75 & 1.01&0.17  & 1.16&0.22  & 0.49&0.14  & 0.58&0.16  \\
\hline
\end{tabular}
}
\end{table*}

$\,\!$\indent We now test which population could be the contaminant in the previous subsections. We start with AGB stars using the catalog of \citet{SuhHong17}
for oxygen-rich (O-AGB) and carbon rich (C-AGB) stars and we add the 
the S stars from \citet{SuhKwon11}. We only select 
objects with 2MASS photometry with quality flag AAA and we divide the samples into two regions: the inner Galactic plane (IGP, defined as the region with 
$|l|\,\le 90^{\rm o}$ and $|b|\,\le 5^{\rm o}$ and composed mostly of high-extinction Galactic disc and bulge stars),
where the sample should experience significant extinction, and the rest of the sky, where NIR extinction should 
be small for most of the sample. Results are plotted in Fig.~\ref{2MASS_cc_AGB} and some relevant statistics are given in Table~\ref{results_AGB} (sample size,
fraction in IGP, $J$ magnitude, and NIR colours). In the bottom panel of Fig.~\ref{2MASS_cc_AGB} the effect of extinction is clearly seen. 
{Blue symbols (non-IGP, mostly low extinction)} 
form a relatively narrow diagonal sequence, 
{with C-AGB stars (triangles)} 
on average being (intrinsically) redder than 
{O-AGB stars (squares). Red symbols (IGP)} 
are located mostly to the right of the blue ones, the expected effect if they follow an extinction trajectory quasi-parallel to that of red giants.

S 
stars are too scarce and too blue to be a significant contaminant. C-AGB stars have more promising numbers but have two things against them: (a) even 
though they are intrinsically redder, their intrinsic $H-K$ values are too red for a given $J-H$ so their extinction trajectories fall above the one for
$\alpha=1$ for RC stars; and (b) most are not in the inner Galactic plane (Table~\ref{results_AGB}). This leaves us with O-AGB stars. They are the most numerous 
group, most of them are in the IGP (Table~\ref{results_AGB}), and their bluer intrinsic colours allow their extinction trajectories to fall in the 
region between the black lines in Fig.~\ref{2MASS_cc_AGB}. In graphical terms, most points below the $\alpha=1$ line for RC stars in the bottom panel of 
Fig.~\ref{2MASS_cc_AGB} are 
{red (IGP) squares (O-AGB).} 
Also, O-AGB stars are known to be much more abundant than C-AGB stars in the bulge and that explains the result from the
previous subsection, in which we found more contaminants in the central (bulge+disc) that in the lateral (disc) regions.
Furthermore, the \citet{SuhHong17} sample is obviously just the tip of the iceberg of all O-AGB stars in the Galaxy 
(as many are missing in regions with very high extinction or that are highly crowded) and their $J$ magnitudes fall in the middle of 
the range of our global 2MASS sample, so many more other stars of the same type must be hidden in our sample there. Therefore, O-AGB stars appear to be a good 
candidate for the dominant contaminant population.

\begin{figure}[h!]
\centerline{\includegraphics[width=\linewidth]{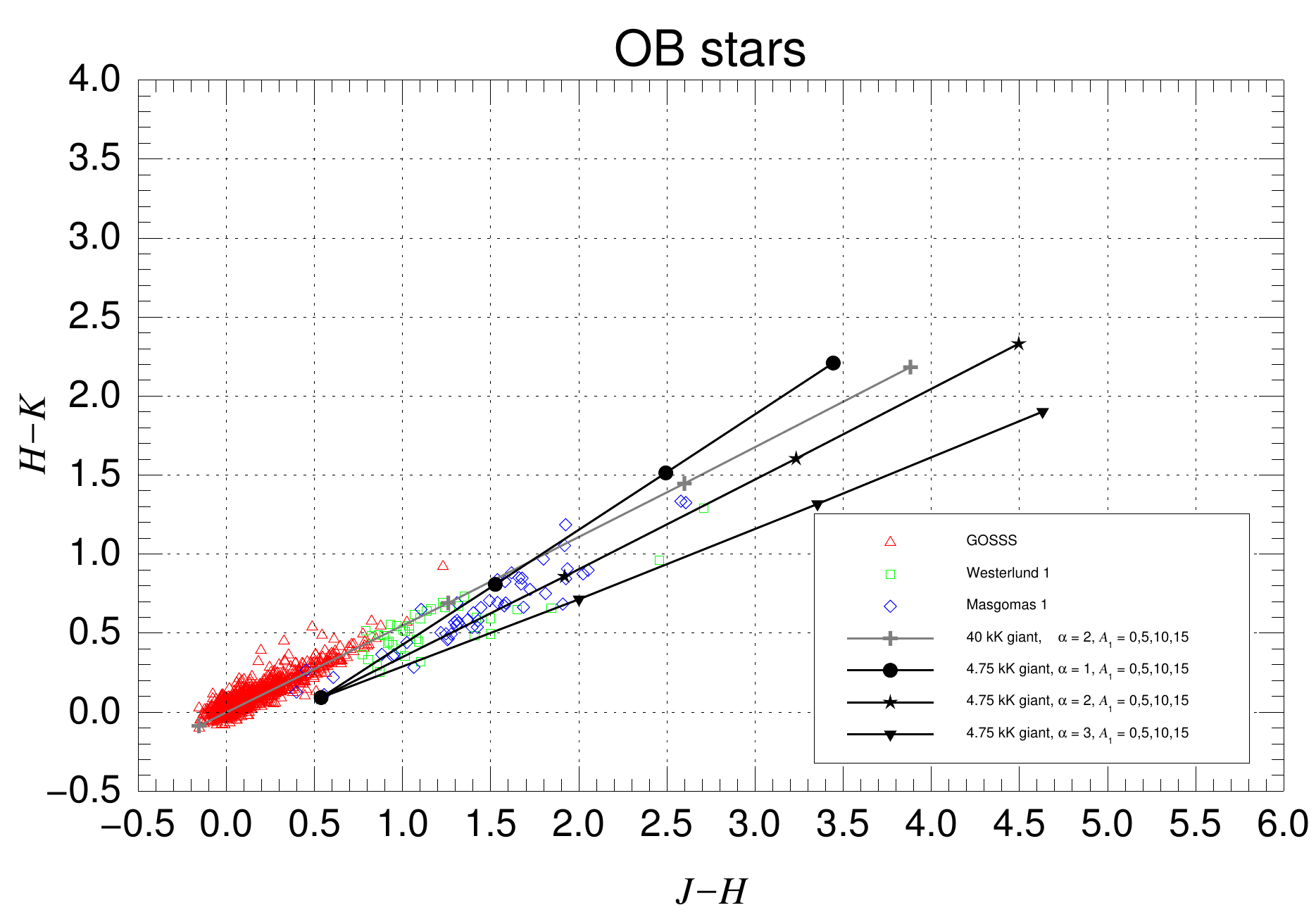}}
\caption{Colour-colour plot for GOSSS-selected O-type stars and for sources in our sample in the regions centred on the Westerlund~1 and Masgomas~1 young
         clusters. The gray line with symbols show the extinction trajectory for $\alpha$~=~2 and for the $A_1$ range between 0~and~15~mag for a 40~kK giant.
         The three black lines with symbols show the extinction trajectories for $\alpha$~=~1,~2,~and~3 and for the $A_1$ range between 0~and~15~mag for a 
         4.75~kK giant.}
\label{2MASS_cc_OB}
\end{figure}

We now attempt the alternative option of extinguished OB stars. We start with O stars with accurate optical spectral classifications from the Galactic O-Star
Spectroscopic Survey (GOSSS, \citealt{Maizetal11}), as shown in Fig.~\ref{2MASS_cc_OB}. Those objects indeed follow the expected extinction trajectory for hot
stars quite tightly but they fall short of reaching the region of interest for contamination because their extinction is not high enough. We also include 
in Fig.~\ref{2MASS_cc_OB} two regions centred on dense young clusters with high extinction, Westerlund~1 and Masgomas~1. In those cases the selection is made 
only by position (not by spectral classification) using our sample (2MASS AAA stars with uncertainties lower than 0.05~mag). There we see a few stars
following the extinction trajectory for hot stars and with extinctions higher than those of the GOSSS sample but in most cases without reaching colours as red as 
what we would need to make a large contribution to the contaminant population. Furthermore, for Westerlund~1 and Masgomas~1 we also see other stars that are 
likely extinguished late-type stars (either cluster supergiants or field red giants). We have also tried other similar clusters located at distances of a few kpc
and the results are even worse in the sense of having a larger proportion of extinguished late-type stars. \citet{Comeetal02} found a similar result in their
analysis of Cyg~OB2. If we go to young clusters with even higher
extinctions, such as the Arches or Quintuplet clusters close to the Galactic Center, something else happens: even though their extinctions are high enough to
yield the right colours, there are just a few or no stars in the 2MASS sample because their uncertainties are larger than 0.05~mag (especially in $J$) or they are 
not detected at all. If this happens in the most favorable conditions (known clusters with high concentrations of extinguished OB stars), finding large numbers of
such stars in the general population should be even more difficult. Therefore, we conclude that highly extinguished OB stars may be only a small contaminant. It 
is not that they do not exist, what happens is that they are not present in the sample because they are too dim to be included in it. 

{Finally, we consider the possibility that some contaminants may be artificial i.e. that they are actually blended sources. One indication that this 
effect may be a real contribution is that \fex\ is larger in the central region of the Galactic Plane (where the source density is higher) than in the lateral region. 
We have visually inspected some crowded regions and compared the 2MASS identifications of some of our sources with those of other surveys with better spatial resolutions 
such as the UKIDSS Galactic Plane Survey \citep{Lucaetal08} or VVV \citep{Minnetal10} and we have indeed found some cases where a given 2MASS source corresponds to two or
three sources in them.} 

\subsection{Detailed spatial distribution}

\begin{figure}
\centerline{\includegraphics[width=\linewidth]{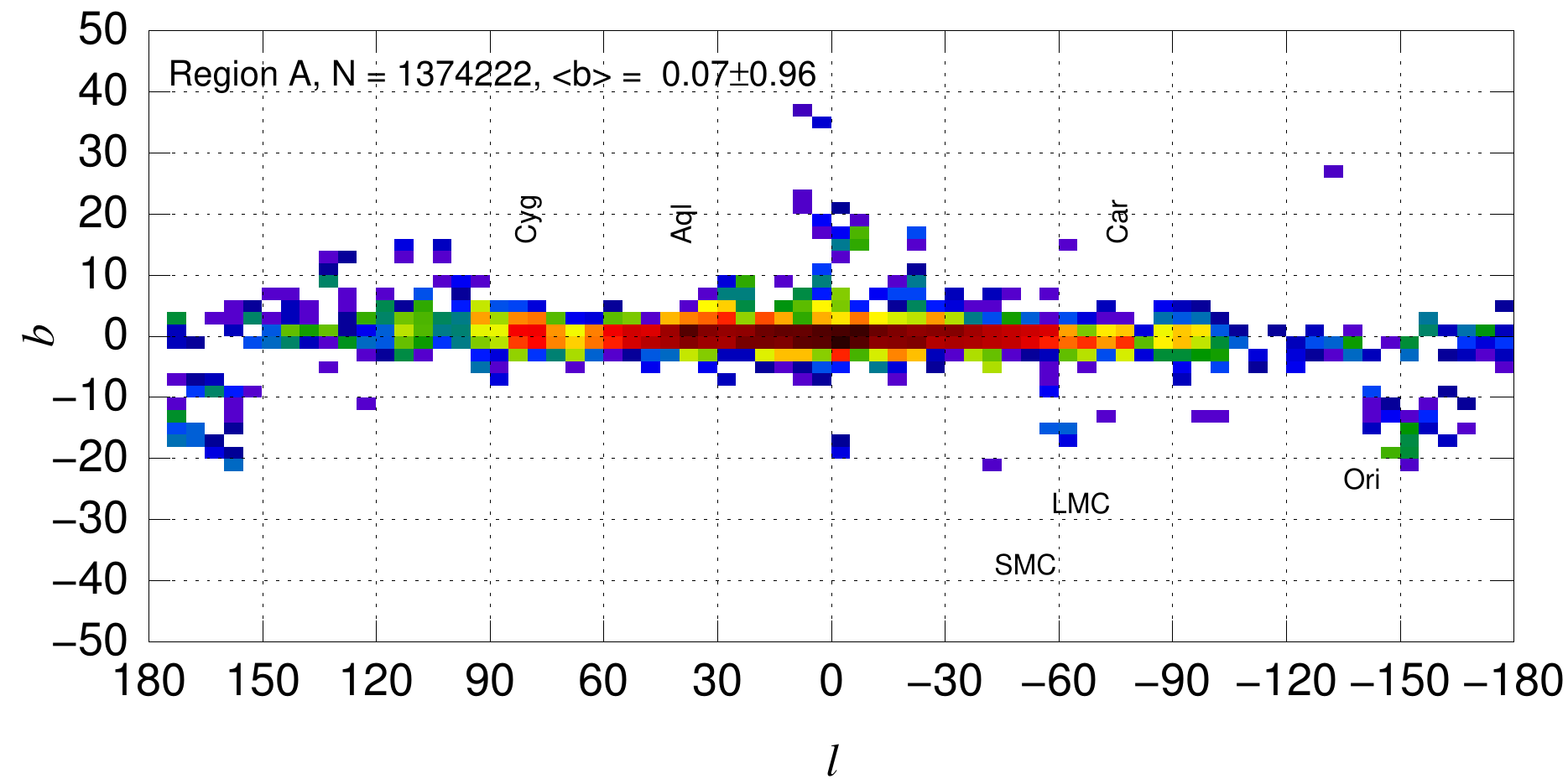}}
\centerline{\includegraphics[width=\linewidth]{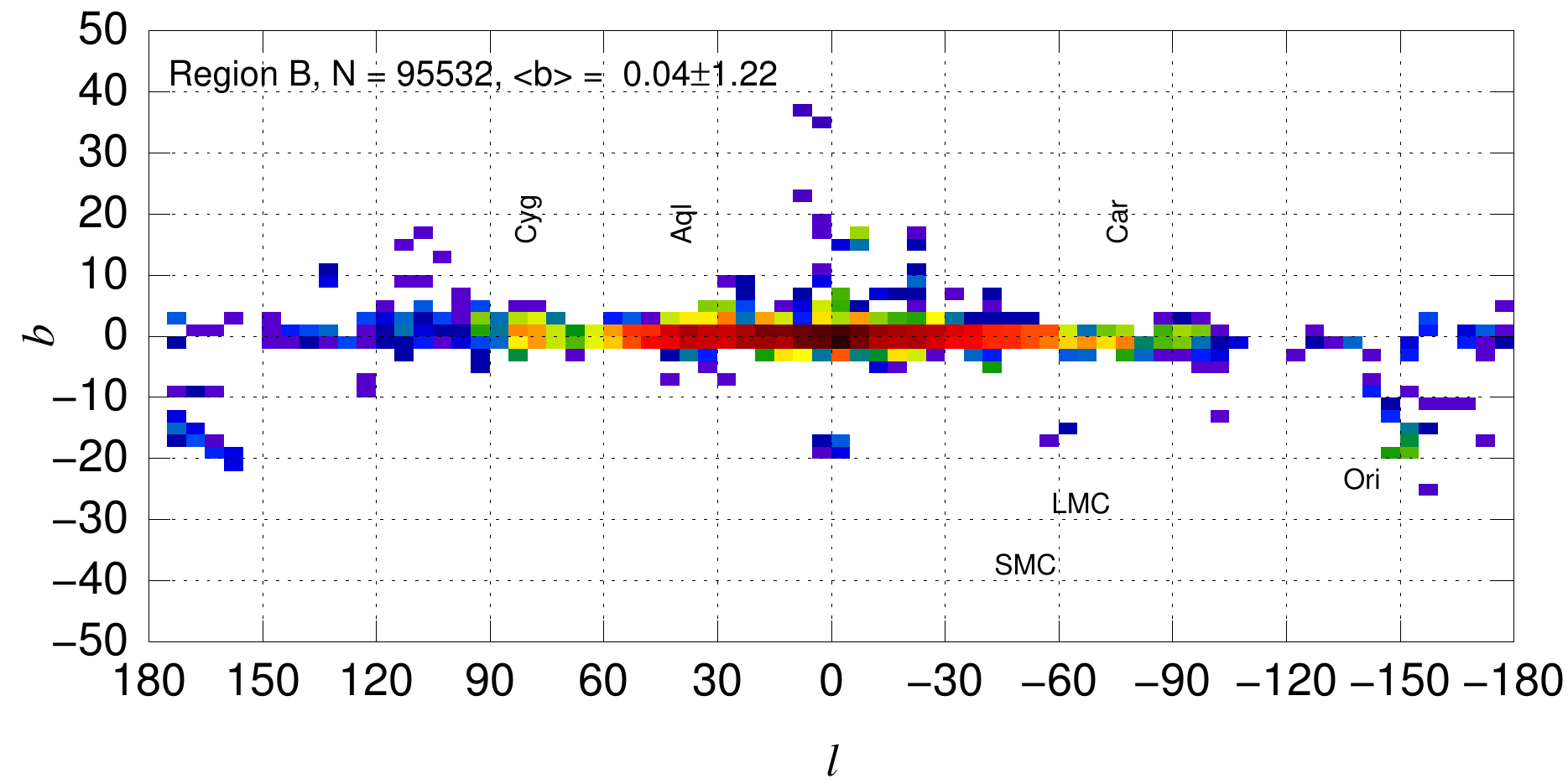}}
\centerline{\includegraphics[width=\linewidth]{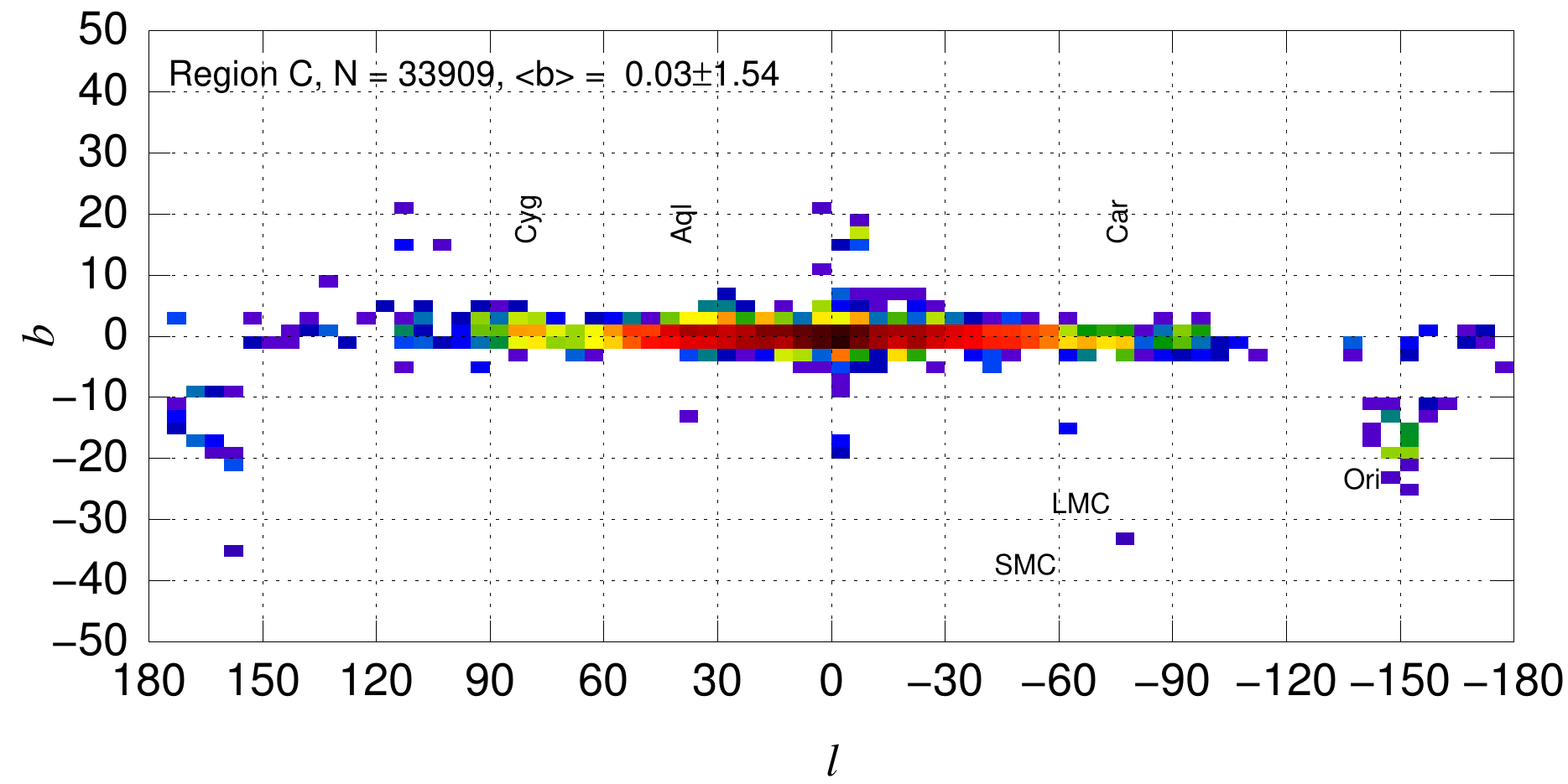}}
\centerline{\includegraphics[width=\linewidth]{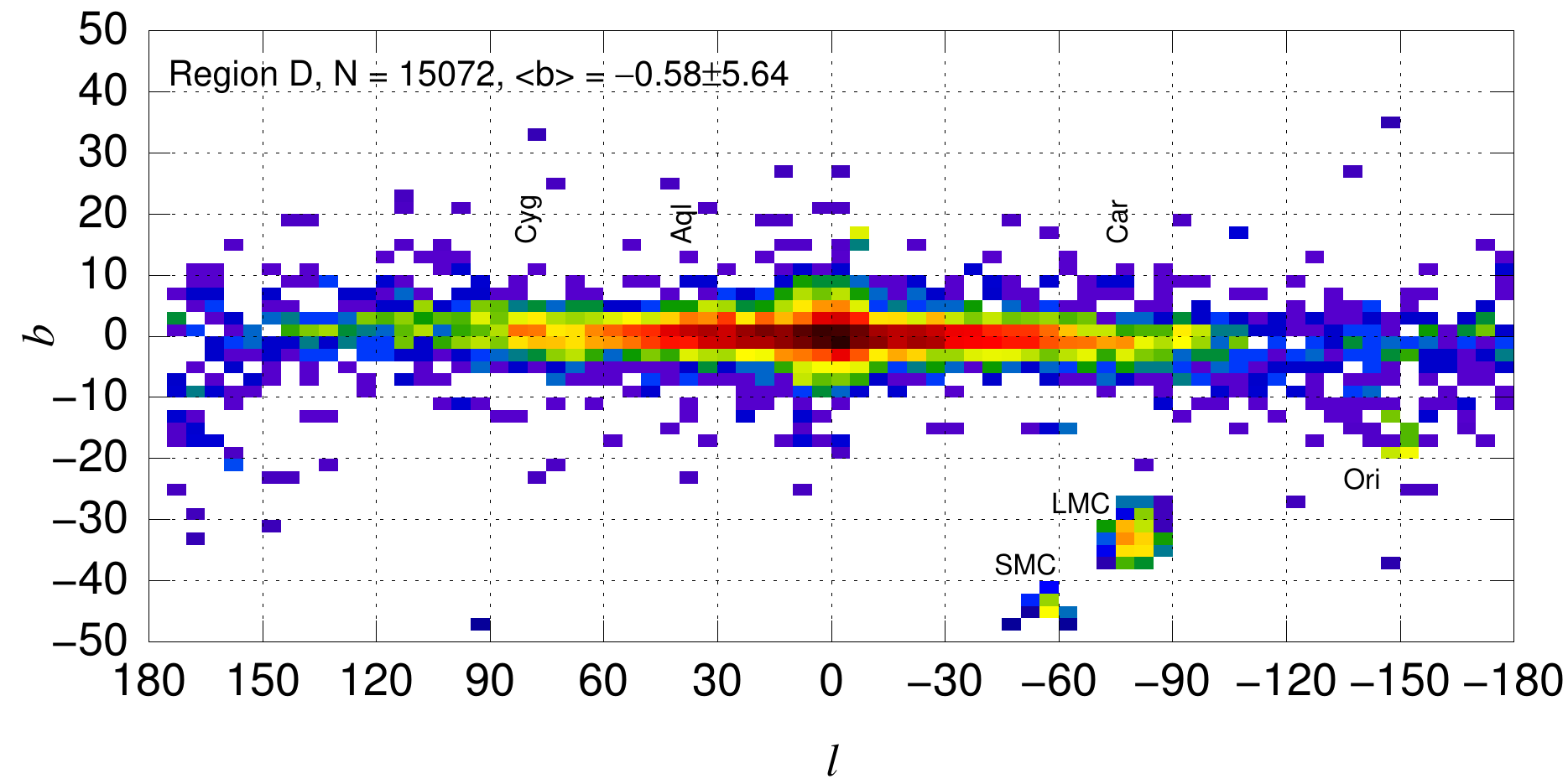}}
\caption{Plane-of-the-sky distribution (log scale) for the four samples in the colour-colour plane defined in the text. The number of stars in each sample and the
         average Galactic latitude and dispersion are given at the top of each panel. Some relevant regions of the sky are labelled.}
\label{region_density}
\end{figure}

\begin{figure}
\centerline{\includegraphics[width=\linewidth]{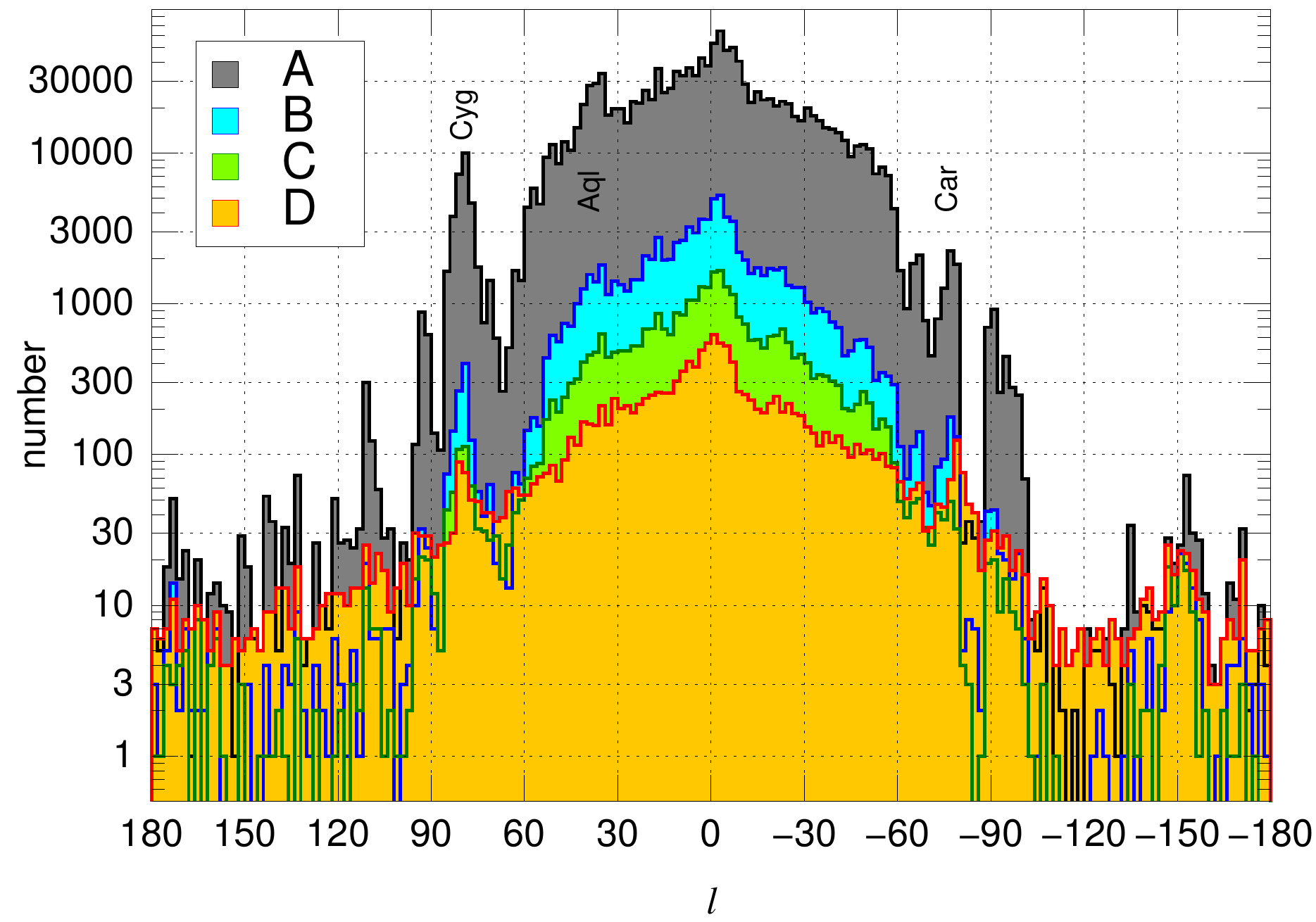}}
\caption{Galactic longitude distribution for the four samples in the colour-colour plane defined in the text. Some relevant regions of the sky are labelled.}
\label{region_lon}
\end{figure}

$\,\!$\indent To confirm that the contaminant population is composed mainly of O-AGB stars we can study the detailed spatial distribution as a function of location in
the $H-K$ vs. $J-H$ plane. We select only those stars with $J-H > 1.8$ and use as dividing lines the extinction trajectories for $\alpha$ of 1.8, 1.5, and 1.0
(all for the standard case of a 4.75~kK~giant) to define four regions in the colour-colour plane: A ($1.8 \,\le \alpha$, expected to be mostly RC stars), B
($1.5 \,\le \alpha < 1.8$, expected to be a mixture of RC and extinguished O-AGB stars), C ($1.0 \,\le \alpha < 1.5$, expected to be mostly extinguished O-AGB 
stars), and D ($\alpha < 1.0$, expected to be AGB stars of diverse type and extinction). The distribution in the plane of the sky of the samples 
associated with each of those four regions is given in Fig.~\ref{region_density} and the histograms of Galactic longitude distributions are given in 
Fig.~\ref{region_lon}, where the sample numbers and basic properties of the distribution in Galactic latitude are also provided.

As expected, the sample numbers monotonically decrease from A to D while the dispersion in $b$ runs in the opposite direction. The latter quantity increases
slowly from A to C while remaining small and then is significantly larger for D. This is a sign that the first three samples are dominated by highly extinguished
sources in the plane of the sky (with the average amount of extinction and hence distance decreasing from A to C) and that the last sample includes a large number of
low-extinction objects. This is seen not only on the increased thickness of the distribution around the Galactic plane in the bottom panel of
Fig.~\ref{region_density} but also in the clear detection of the low-extinction AGB stars (mostly C-rich) in the Magellanic Clouds.

Regarding the distribution in Galactic longitude, the first three samples are heavily concentrated in the inner two quadrants while that of the last sample is 
less concentrated in the same way. Another important difference is that spatial uniformity increases from A to D. The first sample is very patchy as a result of
extinction and that is shown by the existence of significant peaks in the top panel of Fig.~\ref{region_density} and in the black histogram of Fig.~\ref{region_lon}. 
Those peaks become weaker in samples B and C and are hard to identify in D, 
another sign of the decrease in average extinction as we go from the first to the last sample. The three most relevant peaks (in Carina, Aquila, and Cygnus) are 
off-centre sightlines of high extinction with corresponding larger numbers of obscured RC stars, an additional reason why the fourth panel of 
Fig.~\ref{alpha_density} has a lower fraction of contaminants than the third panel. In summary, the analysis of the spatial distribution of the four samples
confirms that O-AGB stars are the dominant contaminant population, which is more abundant in sightlines that include the bulge that in those that do not.
{However, as we previously mentioned, there must also be some contribution from blending in crowded regions.} 

\subsection{Using Gaia to select RC stars}

\begin{figure}
\centerline{\includegraphics[width=\linewidth]{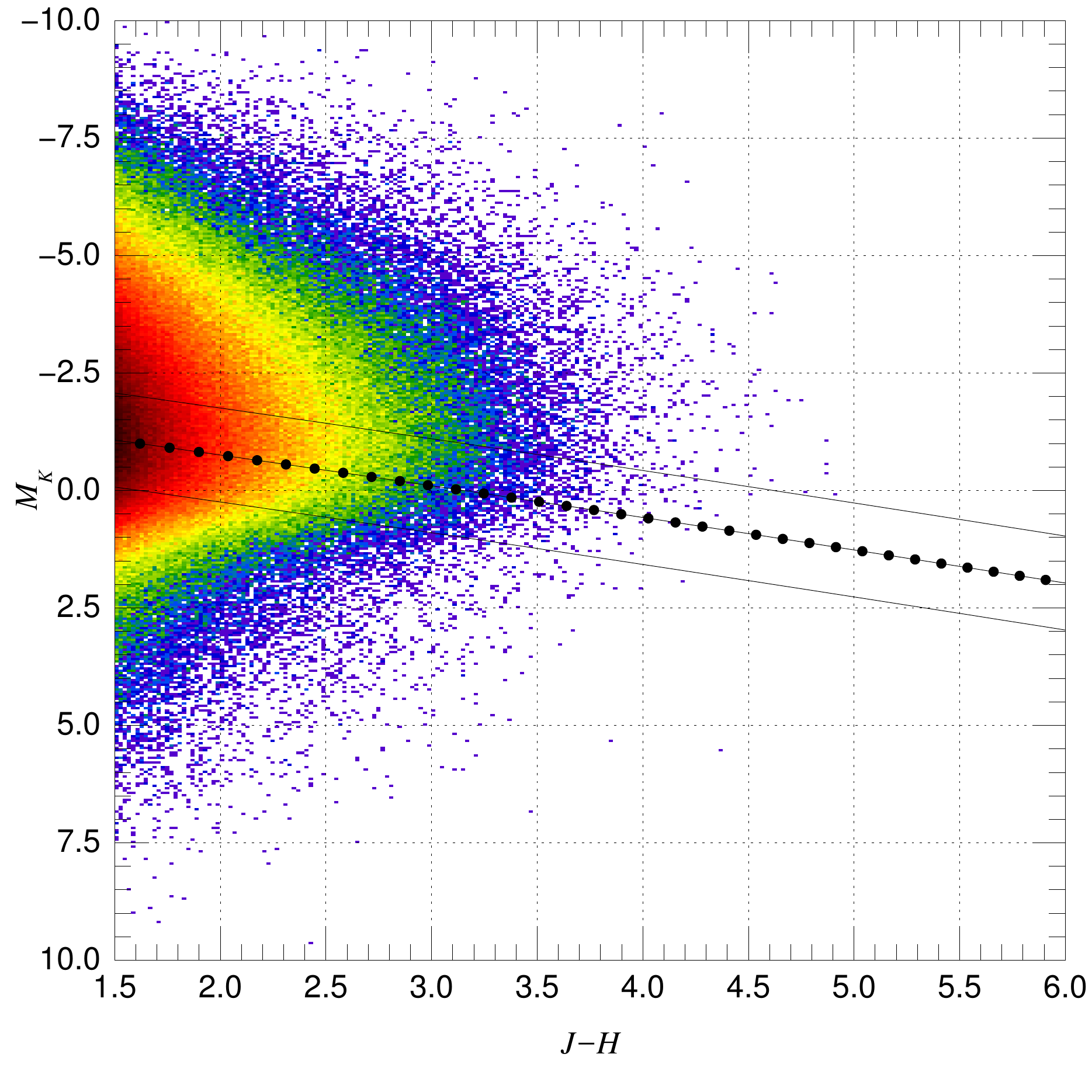}}
\caption{$J-H$ vs. $M_K$ logarithmic density diagram for the {\it Gaia} sample defined in the text. The line with circles shows the extinction trajectory with 
         $\alpha = 2.25$ for RC stars and the other two lines show the limits used to select the final sample.}
\label{CAMD}
\end{figure}

$\,\!$\indent In order to confirm our results, we select a final additional sample by cross-matching 
{(using a 1\arcsec\ radius)} 
our original one with the \citet{Bailetal18} {\it Gaia}~DR2 
sample with distances. After performing the cross-match we select stars with (a) $J-H\,\ge 1.5$, 
{(b) relative distance uncertainties [defined as $0.5(r_{\rm hi}-r_{\rm lo})/r_{\rm est}$, see \citealt{Bailetal18} for definitions] of 50\% or less,} 
and (c) located in a 
2~mag~wide band around the expected extinction trajectory for $\alpha = 2.25$ for RC stars in the resulting $J-H$~vs.~$M_K$ diagram (Fig.~\ref{CAMD}, note that
selecting a similar but different value of $\alpha$ does not introduce significant changes). For the zero-extinction $M_K$ for RC stars we use $-1.622$ 
\citep{ChanBovy20}. The second selection criterion is relatively broad, as most 
extinguished RC stars in {\it Gaia}~DR2 have large parallax uncertainties and selecting a more severe restriction would reduce the sample significantly. This 
means that our {\it Gaia} sample should have a small (but reduced) fraction of contaminants.

\begin{figure}
\centerline{\includegraphics[width=\linewidth]{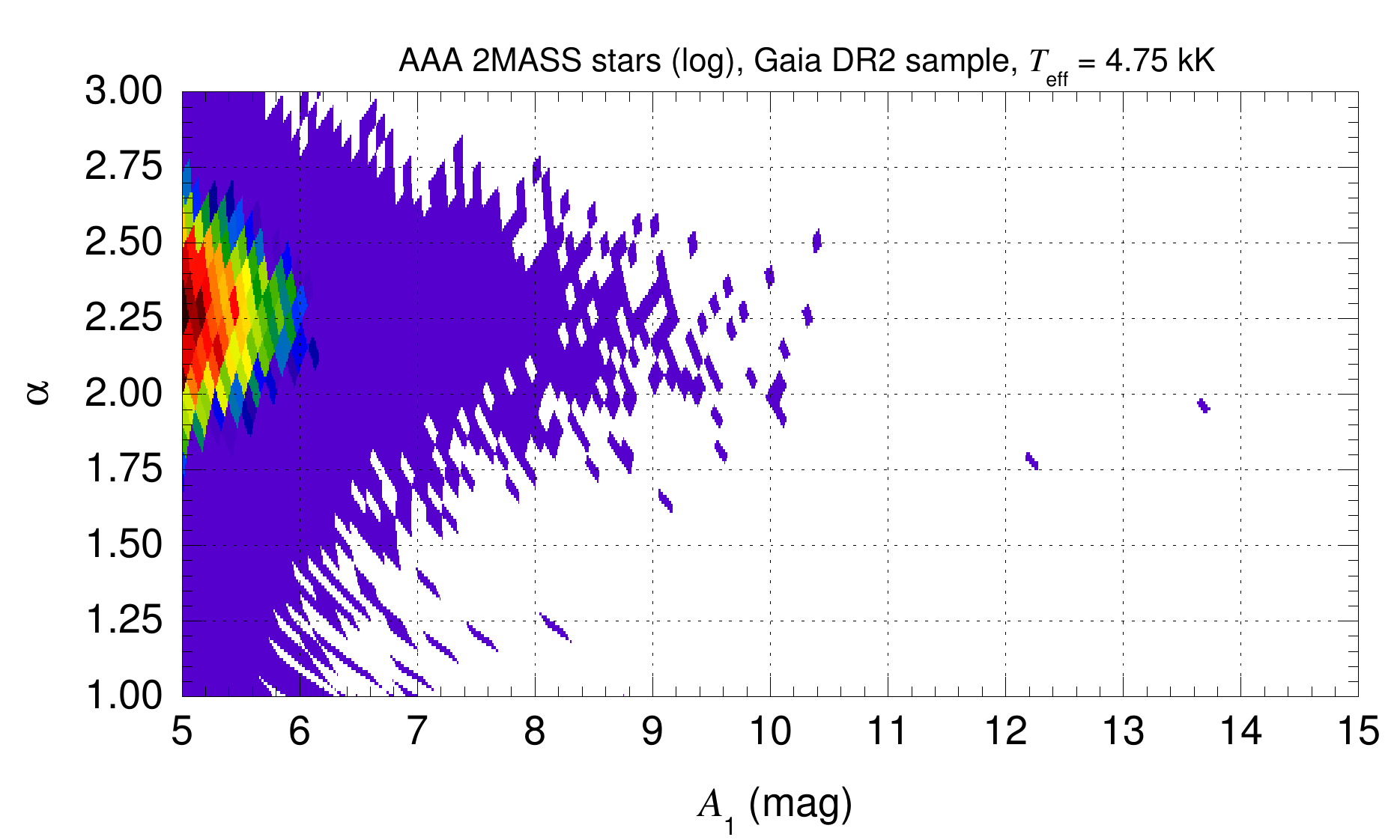}}
\vspace{-2mm}
\centerline{\includegraphics[width=\linewidth]{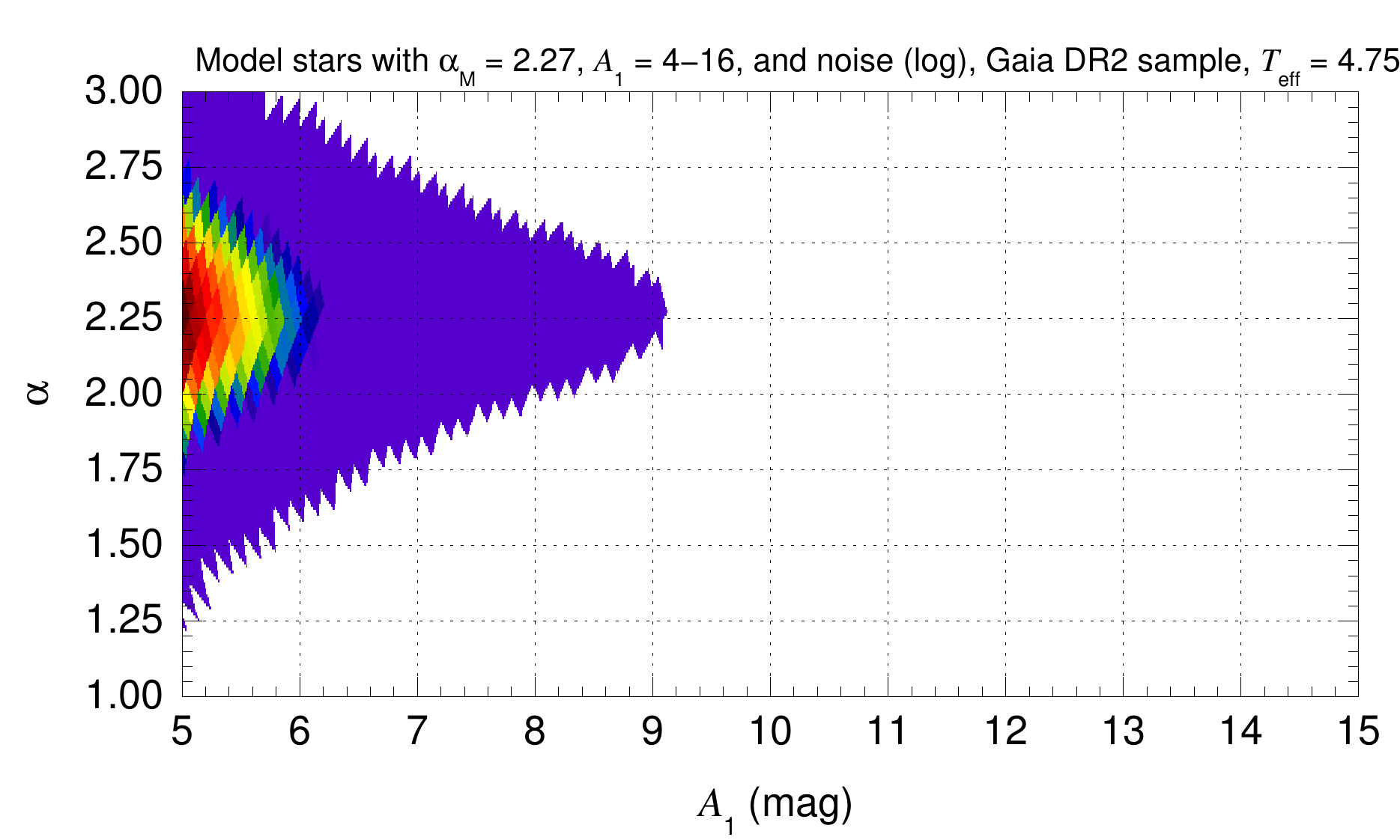}}
\vspace{-2mm}
\centerline{\includegraphics[width=\linewidth]{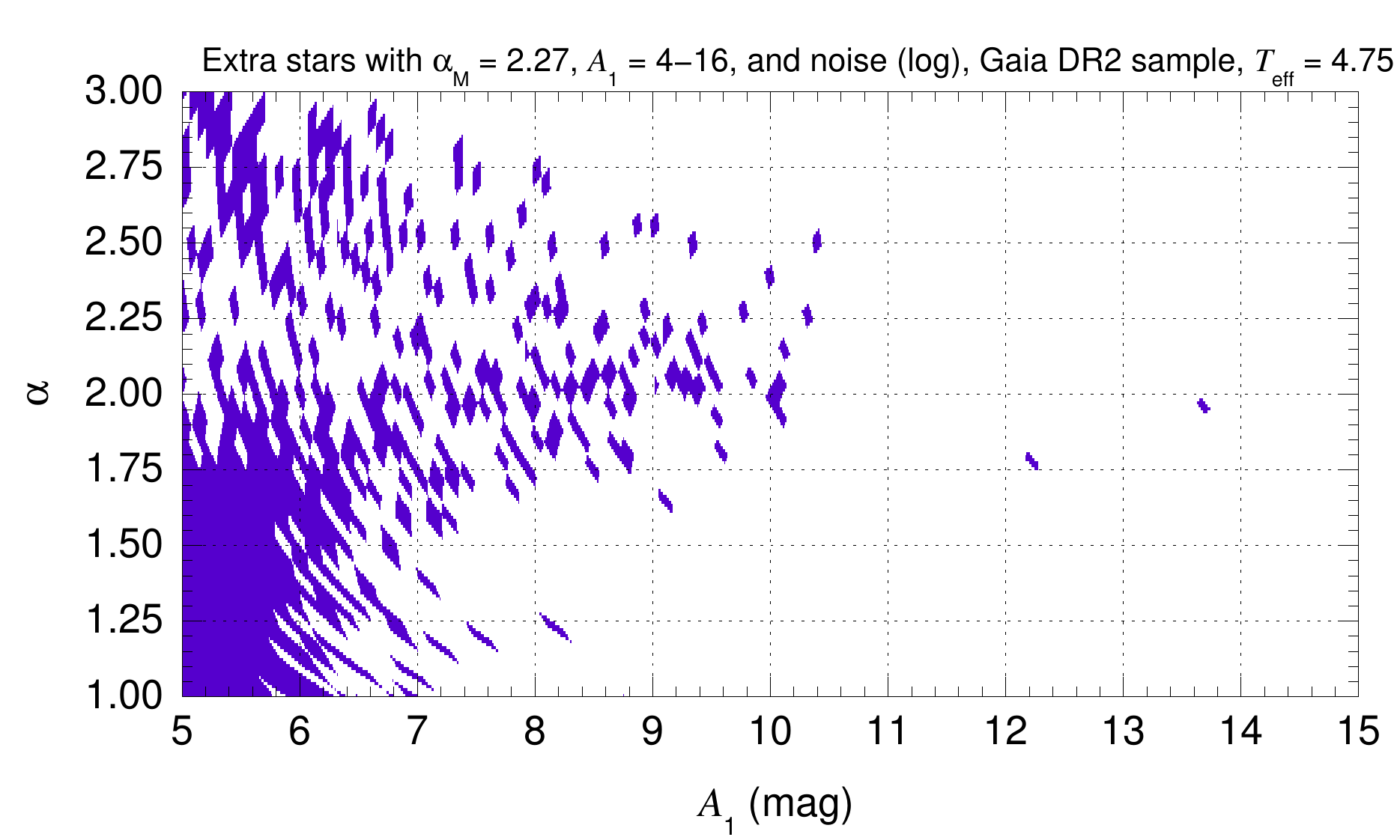}}
\vspace{-2mm}
\centerline{\includegraphics[width=\linewidth]{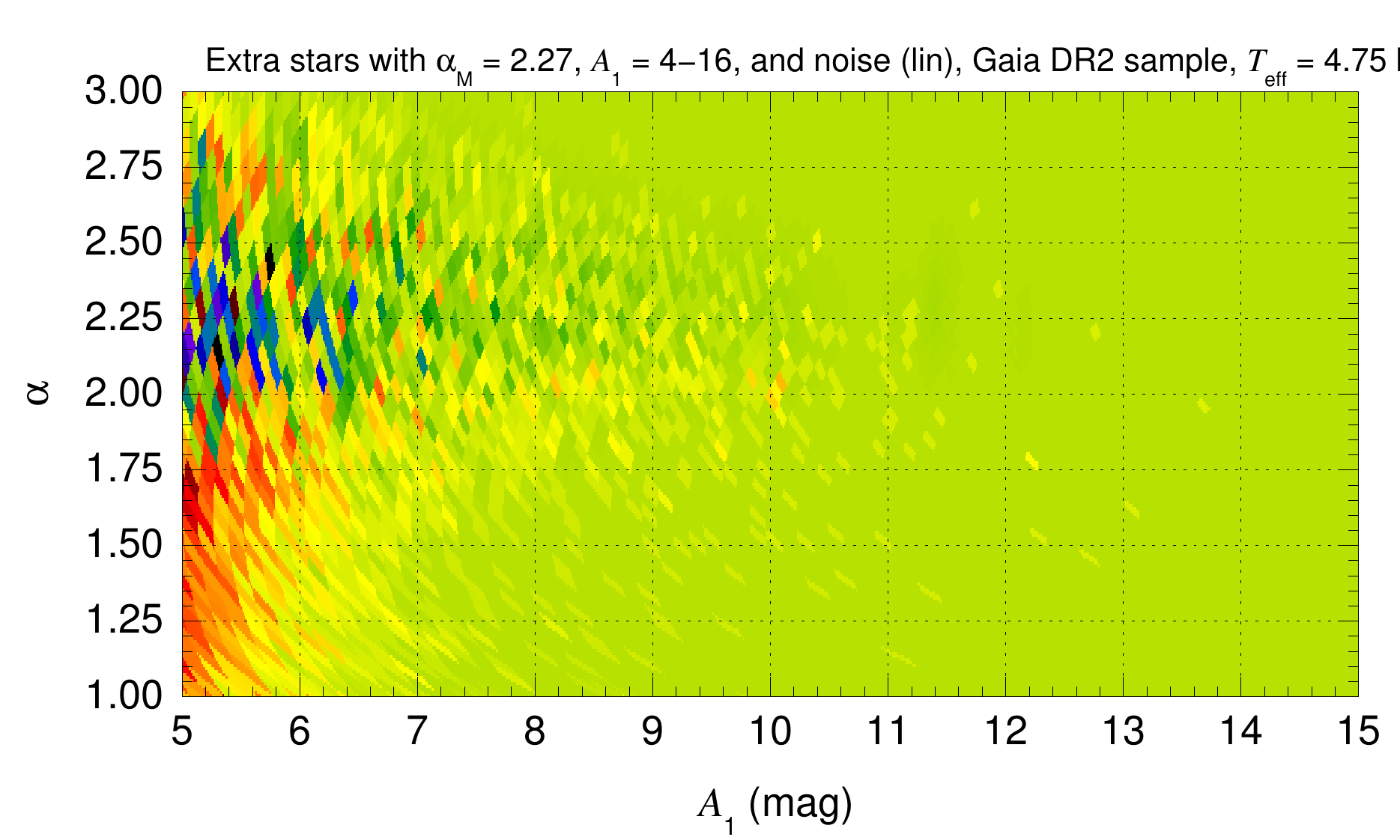}}
\caption{Same as Fig.~\ref{2MASS_par_05_4750} for the {\it} Gaia sample.}
\label{2MASS_par_final_4750}
\end{figure}

The results for the {\it Gaia} sample are given in Table~\ref{results} and Figs.~\ref{alpha_density}~and~\ref{2MASS_par_final_4750}. We obtain 
a value of \alphaM\ (2.27) just slightly higher than the previous ones and a lower contamination fraction than for our previous attempts. The wing is only 
weakly seen for the lowest values of $A_1$ in Fig.~\ref{2MASS_par_final_4750}. Therefore, we conclude that Galactic RC stars are extinguished with an
$\alpha$ of 2.27, with a systematic uncertainty of a few hundredths of a magnitude determined by modelling and the presence of contaminants and a real spread 
that cannot be larger than that because otherwise we would observe a larger width of the distribution in the bottom panel of Fig.~\ref{alpha_density}. 
{\bf The data show no indication of significant variations in the NIR extinction law as a function of sightline or amount of extinction.} 
{This result is consistent with the finding by \citet{Noguetal18b} that different regions in the inner bulge have similar values of $\alpha$ 
compatible with the one that we calculate here even though their amounts of extinction are very different. 
An $\alpha$ value around 2.27 should be valid for the conditions described in this paper i.e. red giants affected by typical Galactic plane extinction. It is possible
that under other circumstances, such as those present in molecular clouds associated with star-forming regions, a different value of $\alpha$ is found. Indeed, in
paper II (submitted to MNRAS) we have found out precisely such a discrepancy in the extinction law of molecular clouds with respect to the typical Galactic extinction
in the 7700~\AA\ region.} 

\section{The past and the future}

$\,\!$\indent There is an extensive literature on NIR extinction, with many authors calculating a value of $\alpha$ and a large diversity of results. The measured
values have increased over the years from the pioneer measurements of 1.61 by \citet{RiekLebo85} and 1.77 by \citet{Drai89} to the one of 1.99 of 
\citet{Nishetal09} and the higher, more 
recent ones of \citet{Fritetal11}, $2.11\pm 0.06$, \citet{Alonetal17a}, $2.47\pm0.11$, and \citet{Noguetal18a}, $2.30\pm0.08$, among others. Considering that in 
some cases the samples are similar or even overlap, it is logical to suspect that there are calibration or methodological problems at work. In the case of 
\citet{RiekLebo85} one culprit is clear, the inclusion of Cyg~OB2-12 in the example, a B-type variable hypergiant with a strong wind \citep{Salaetal15,Nazeetal19}
and a significant IR excess. In others, however, we suspect that at least part of the problem lies in the linearization of extinction calculations, as most 
previous NIR papers do not appear to take the non-linear behavior described in the appendix here into account (there are some exceptions, of course, e.g. 
\citealt{SteaHoar09,WangChen19}). The evolution from low to high values of $\alpha$ since the 1980s also points towards the correction of systematic effects in the 
analysis.

One previous paper did an analysis similar to the one here, \citet{SteaHoar09}. They used a combination of UKIDSS and 2MASS photometry and an explicit treatment of
non-linear effects to conclude that $\alpha = 2.14^{+0.04}_{-0.05}$ with little variation from sightline to sightline. Our sample is much larger and more complete, as
we selected sources from the whole sky following photometric quality criteria instead of concentrating on several specific regions. We reach the same conclusion with
respect to the uniformity of $\alpha$ but our value is somewhat higher. We suspect the difference arises due to the presence of contaminants in their sample, which,
as we have seen here, tends to drive the value of $\alpha$ down.


There is another important reason why different methods produce different values of $\alpha$: the NIR extinction law likely does not follow a power law. Different
papers point in the direction that the extinction law becomes flatter as $\lambda$ increases. \citet{Noguetal19} measure $\alpha = 2.43\pm0.10$ in the range of
the $JH$ bands but $\alpha = 2.23\pm0.03$ in the range of the $HK$ bands. \citet{Hoseetal18} use additional NIR filters to detect that a power law cannot
consistently be fitted to all of them. Also, when one moves to wavelengths longer than 2.5~$\mu$m the extinction law is flatter \citep{Fritetal11}, so it makes 
sense that there is a transition that affects the $K$ band already. Therefore, any result on $\alpha$ (such as the one in this paper) should be taken as an
approximation to reality that requires further improvement. A consequence of this is that the $A_1$ values calculated from $JHK$ magnitudes using a power 
law are likely underestimates, as the extinction law at 1~$\mu$m is higher than what those bands predict. Beyond 1~$\mu$m the situation changes again, as the
extinction law becomes flatter in the optical and clearly dependent on the sightline \citep{Fritetal11,Maizetal14a,Meinetal18}.

Our plans for the future involve several lines of work:

\begin{itemize}
 \item Combine 2MASS with WISE photometry to extend our analysis to longer wavelengths. In those cases with high extinctions we will also include {\it Gaia} 
       $G$ photometry, thanks to its excellent filter characterization \citep{MaizWeil18}, to tie in the NIR extinction with its value in the 0.9-1.0~$\mu$m
       range.
 \item Study the optical extinction law combining {\it Gaia} data with the GALANTE photometric survey \citep{Maizetal19c,LorGetal19} and other deep surveys of the
       Galactic plane.
 \item There is a limit to what can be studied about the extinction law with photometry. The right way to do it is with spectrophotometry and for that purpose we
       are obtaining ground-based and HST spectroscopic data in the UV, optical, and NIR.
\end{itemize}

\section*{Acknowledgements}
$\,\!$\indent {We thank an anonymous referee for helping us improve the paper, especially regarding the role of blended sources as contaminants.} 
J.M.A. and M.P.G. acknowledge support from the Spanish Government Ministerio de Ciencia through grant PGC2018-095\,049-B-C22. 
R.H.B. acknowledges support from DIDULS Project 18\,143 and the ESAC Faculty Visitor Program.
F.N.-L. acknowledges funding by the Deutsche Forschungsgemeinschaft (DFG, German Research Foundation) - project ID 138713538 - SFB 881 
(``The Milky Way System'', subproject B8).

\section*{Data availability} 
$\,\!$\indent This publication makes use of data products from the Two Micron All Sky Survey, which is a joint project of the University of Massachusetts and the Infrared 
Processing and Analysis Center/California Institute of Technology, funded by the National Aeronautics and Space Administration and the National Science Foundation,
where the word ``National'' refers to the United States of America.
It also makes use of data from the European Space Agency (ESA) mission {\it Gaia} (\url{https://www.cosmos.esa.int/gaia}), processed by the 
{\it Gaia} Data Processing and Analysis Consortium (DPAC, \url{https://www.cosmos.esa.int/web/gaia/dpac/consortium}). Funding for the DPAC has been 
provided by national institutions, in particular the institutions participating in the {\it Gaia} Multilateral Agreement. 
The derived data generated in this research will be shared on reasonable request to the corresponding author.

\bibliographystyle{mnras} 
\bibliography{general} 

\begin{thebibliography}{}
\makeatletter
\relax
\def\mn@urlcharsother{\let\do\@makeother \do\$\do\&\do\#\do\^\do\_\do\%\do\~}
\def\mn@doi{\begingroup\mn@urlcharsother \@ifnextchar [ {\mn@doi@}
  {\mn@doi@[]}}
\def\mn@doi@[#1]#2{\def\@tempa{#1}\ifx\@tempa\@empty \href
  {http://dx.doi.org/#2} {doi:#2}\else \href {http://dx.doi.org/#2} {#1}\fi
  \endgroup}
\def\mn@eprint#1#2{\mn@eprint@#1:#2::\@nil}
\def\mn@eprint@arXiv#1{\href {http://arxiv.org/abs/#1} {{\tt arXiv:#1}}}
\def\mn@eprint@dblp#1{\href {http://dblp.uni-trier.de/rec/bibtex/#1.xml}
  {dblp:#1}}
\def\mn@eprint@#1:#2:#3:#4\@nil{\def\@tempa {#1}\def\@tempb {#2}\def\@tempc
  {#3}\ifx \@tempc \@empty \let \@tempc \@tempb \let \@tempb \@tempa \fi \ifx
  \@tempb \@empty \def\@tempb {arXiv}\fi \@ifundefined
  {mn@eprint@\@tempb}{\@tempb:\@tempc}{\expandafter \expandafter \csname
  mn@eprint@\@tempb\endcsname \expandafter{\@tempc}}}

\bibitem[\protect\citeauthoryear{Alonso-Garc{\'{\i}}a
  et~al.,}{Alonso-Garc{\'{\i}}a et~al.}{2017}]{Alonetal17a}
Alonso-Garc{\'{\i}}a J.,  et~al., 2017, \mn@doi [ApJL]
  {10.3847/2041-8213/aa92c3}, \href
  {http://adsabs.harvard.edu/abs/2017ApJ...849L..13A} {849, L13}

\bibitem[\protect\citeauthoryear{Bailer-Jones, Rybizki, Fouesneau, Mantelet  \&
  Andrae}{Bailer-Jones et~al.}{2018}]{Bailetal18}
Bailer-Jones C.~A.~L.,  Rybizki J.,  Fouesneau M.,  Mantelet G.,   Andrae R.,
  2018, \mn@doi [AJ] {10.3847/1538-3881/aacb21}, \href
  {http://adsabs.harvard.edu/abs/2018AJ....156...58B} {156, 58}

\bibitem[\protect\citeauthoryear{Blanco}{Blanco}{1956}]{Blan56}
Blanco V.~M.,  1956, \mn@doi [ApJ] {10.1086/146131}, \href
  {http://adsabs.harvard.edu/abs/1956ApJ...123...64B} {123, 64}

\bibitem[\protect\citeauthoryear{Blanco}{Blanco}{1957}]{Blan57}
Blanco V.~M.,  1957, \mn@doi [ApJ] {10.1086/146294}, \href
  {http://adsabs.harvard.edu/abs/1957ApJ...125..209B} {125, 209}

\bibitem[\protect\citeauthoryear{Bovy et~al.,}{Bovy et~al.}{2014}]{Bovyetal14}
Bovy J.,  et~al., 2014, \mn@doi [ApJ] {10.1088/0004-637X/790/2/127}, \href
  {https://ui.adsabs.harvard.edu/abs/2014ApJ...790..127B} {790, 127}

\bibitem[\protect\citeauthoryear{Chan \& Bovy}{Chan \& Bovy}{2020}]{ChanBovy20}
Chan V.~C.,  Bovy J.,  2020, \mn@doi [MNRAS] {10.1093/mnras/staa571}, \href
  {https://ui.adsabs.harvard.edu/abs/2020MNRAS.493.4367C} {493, 4367}

\bibitem[\protect\citeauthoryear{Comer{\'o}n et~al.,}{Comer{\'o}n
  et~al.}{2002}]{Comeetal02}
Comer{\'o}n F.,  et~al., 2002, A\&A, \href
  {http://adsabs.harvard.edu/abs/2002A&A...389..874C} {389, 874}

\bibitem[\protect\citeauthoryear{Draine}{Draine}{1989}]{Drai89}
Draine B.~T.,  1989, in B{\"o}hm-Vitense E.,  ed., Infrared Spectroscopy in
  Astronomy. p.~93

\bibitem[\protect\citeauthoryear{Fritz et~al.,}{Fritz
  et~al.}{2011}]{Fritetal11}
Fritz T.~K.,  et~al., 2011, \mn@doi [ApJ] {10.1088/0004-637X/737/2/73}, \href
  {http://adsabs.harvard.edu/abs/2011ApJ...737...73F} {737, 73}

\bibitem[\protect\citeauthoryear{Girardi}{Girardi}{2016}]{Gira16}
Girardi L.,  2016, \mn@doi [ARA\&A] {10.1146/annurev-astro-081915-023354},
  \href {https://ui.adsabs.harvard.edu/abs/2016ARA&A..54...95G} {54, 95}

\bibitem[\protect\citeauthoryear{Gonz{\'a}lez, Rejkuba, Zoccali, Valenti,
  Minniti, Schultheis, Tobar  \& Chen}{Gonz{\'a}lez et~al.}{2012}]{Gonzetal12}
Gonz{\'a}lez O.~A.,  Rejkuba M.,  Zoccali M.,  Valenti E.,  Minniti D.,
  Schultheis M.,  Tobar R.,   Chen B.,  2012, \mn@doi [A\&A]
  {10.1051/0004-6361/201219222}, \href
  {https://ui.adsabs.harvard.edu/abs/2012A&A...543A..13G} {543, A13}

\bibitem[\protect\citeauthoryear{Gustafsson, Edvardsson, Eriksson,
  Mizuno-Wiedner, J{\o}rgensen  \& Plez}{Gustafsson et~al.}{2003}]{Gustetal03}
Gustafsson B.,  Edvardsson B.,  Eriksson K.,  Mizuno-Wiedner M.,  J{\o}rgensen
  U.~G.,   Plez B.,  2003, in Hubeny I.,  Mihalas D.,   Werner K.,  eds,
  Astronomical Society of the Pacific Conference Series Vol. 288, Stellar
  Atmosphere Modeling. pp 331--+

\bibitem[\protect\citeauthoryear{Hosek Jr. et~al.,}{Hosek
  et~al.}{2018}]{Hoseetal18}
Hosek Jr. M.~W.,  et~al., 2018, \mn@doi [ApJ] {10.3847/1538-4357/aaabbb}, \href
  {http://adsabs.harvard.edu/abs/2018ApJ...855...13H} {855, 13}

\bibitem[\protect\citeauthoryear{Jones \& Hyland}{Jones \&
  Hyland}{1980}]{JoneHyla80}
Jones T.~J.,  Hyland A.~R.,  1980, \mn@doi [MNRAS] {10.1093/mnras/192.3.359},
  \href {https://ui.adsabs.harvard.edu/abs/1980MNRAS.192..359J} {192, 359}

\bibitem[\protect\citeauthoryear{Lanz \& Hubeny}{Lanz \&
  Hubeny}{2003}]{LanzHube03}
Lanz T.,  Hubeny I.,  2003, ApJS, \href
  {http://adsabs.harvard.edu/abs/2003ApJS..146..417L} {146, 417}

\bibitem[\protect\citeauthoryear{Lanz \& Hubeny}{Lanz \&
  Hubeny}{2007}]{LanzHube07}
Lanz T.,  Hubeny I.,  2007, ApJS, \href
  {http://adsabs.harvard.edu/abs/2007ApJS..169...83L} {169, 83}

\bibitem[\protect\citeauthoryear{Lorenzo-Guti{\'e}rrez
  et~al.,}{Lorenzo-Guti{\'e}rrez et~al.}{2019}]{LorGetal19}
Lorenzo-Guti{\'e}rrez A.,  et~al., 2019, \mn@doi [MNRAS]
  {10.1093/mnras/stz842}, \href
  {https://ui.adsabs.harvard.edu/abs/2019MNRAS.486..966L} {486, 966}

\bibitem[\protect\citeauthoryear{Lucas et~al.,}{Lucas
  et~al.}{2008}]{Lucaetal08}
Lucas P.~W.,  et~al., 2008, \mn@doi [MNRAS] {10.1111/j.1365-2966.2008.13924.x},
  \href {https://ui.adsabs.harvard.edu/abs/2008MNRAS.391..136L} {391, 136}

\bibitem[\protect\citeauthoryear{Ma{\'{\i}}z~Apell{\'a}niz}{Ma{\'{\i}}z~Apell{\'a}niz}{2004}]{Maiz04c}
Ma{\'{\i}}z~Apell{\'a}niz J.,  2004, PASP, \href
  {http://adsabs.harvard.edu/abs/2004PASP..116..859M} {116, 859}

\bibitem[\protect\citeauthoryear{Ma{\'{\i}}z~Apell{\'a}niz}{Ma{\'{\i}}z~Apell{\'a}niz}{2005}]{Maiz05b}
Ma{\'{\i}}z~Apell{\'a}niz J.,  2005, PASP, \href
  {http://adsabs.harvard.edu/abs/2005PASP..117..615M} {117, 615}

\bibitem[\protect\citeauthoryear{Ma{\'{\i}}z~Apell{\'a}niz}{Ma{\'{\i}}z~Apell{\'a}niz}{2006}]{Maiz06a}
Ma{\'{\i}}z~Apell{\'a}niz J.,  2006, AJ, \href
  {http://adsabs.harvard.edu/abs/2006AJ....131.1184M} {131, 1184}

\bibitem[\protect\citeauthoryear{Ma{\'{\i}}z~Apell{\'a}niz}{Ma{\'{\i}}z~Apell{\'a}niz}{2007}]{Maiz07a}
Ma{\'{\i}}z~Apell{\'a}niz J.,  2007, in Sterken C.,  ed.,  ASP Conf. Series
  Vol. 364, The Future of Photometric, Spectrophotometric and Polarimetric
  Standardization. p.~227

\bibitem[\protect\citeauthoryear{Ma{\'{\i}}z~Apell{\'a}niz}{Ma{\'{\i}}z~Apell{\'a}niz}{2013a}]{Maiz13b}
Ma{\'{\i}}z~Apell{\'a}niz J.,  2013a, in HSA 7. pp 583--589 (\mn@eprint {arXiv}
  {arXiv:1209.2560})

\bibitem[\protect\citeauthoryear{Ma{\'{\i}}z~Apell{\'a}niz}{Ma{\'{\i}}z~Apell{\'a}niz}{2013b}]{Maiz13a}
Ma{\'{\i}}z~Apell{\'a}niz J.,  2013b, in HSA 7. pp 657--657 (\mn@eprint {arXiv}
  {arXiv:1209.1709})

\bibitem[\protect\citeauthoryear{Ma{\'{\i}}z~Apell{\'a}niz}{Ma{\'{\i}}z~Apell{\'a}niz}{2017}]{Maiz17a}
Ma{\'{\i}}z~Apell{\'a}niz J.,  2017, \mn@doi [A\&A]
  {10.1051/0004-6361/201732167}, \href
  {http://adsabs.harvard.edu/abs/2017A&A...608L...8M} {608, L8}

\bibitem[\protect\citeauthoryear{Ma{\'{\i}}z~Apell{\'a}niz \&
  Barb{\'a}}{Ma{\'{\i}}z~Apell{\'a}niz \& Barb{\'a}}{2018}]{MaizBarb18}
Ma{\'{\i}}z~Apell{\'a}niz J.,  Barb{\'a} R.~H.,  2018, \mn@doi [A\&A]
  {10.1051/0004-6361/201732050}, \href
  {http://adsabs.harvard.edu/abs/2018A&A...613A...9M} {613, A9}

\bibitem[\protect\citeauthoryear{Ma{\'{\i}}z~Apell{\'a}niz \&
  Pantaleoni~Gonz{\'a}lez}{Ma{\'{\i}}z~Apell{\'a}niz \&
  Pantaleoni~Gonz{\'a}lez}{2018}]{MaizPant18}
Ma{\'{\i}}z~Apell{\'a}niz J.,  Pantaleoni~Gonz{\'a}lez M.,  2018, \mn@doi
  [A\&A] {10.1051/0004-6361/201833918}, \href
  {http://adsabs.harvard.edu/abs/2018A&A...616L...7M} {616, L7}

\bibitem[\protect\citeauthoryear{Ma{\'{\i}}z~Apell{\'a}niz \&
  Weiler}{Ma{\'{\i}}z~Apell{\'a}niz \& Weiler}{2018}]{MaizWeil18}
Ma{\'{\i}}z~Apell{\'a}niz J.,  Weiler M.,  2018, \mn@doi [A\&A]
  {10.1051/0004-6361/201834051}, \href
  {http://adsabs.harvard.edu/abs/2018A&A...619A.180M} {619, A180}

\bibitem[\protect\citeauthoryear{Ma{\'{\i}}z~Apell{\'a}niz, Sota, Walborn,
  Alfaro, Barb{\'a}, Morrell, Gamen  \& Arias}{Ma{\'{\i}}z~Apell{\'a}niz
  et~al.}{2011}]{Maizetal11}
Ma{\'{\i}}z~Apell{\'a}niz J.,  Sota A.,  Walborn N.~R.,  Alfaro E.~J.,
  Barb{\'a} R.~H.,  Morrell N.~I.,  Gamen R.~C.,   Arias J.~I.,  2011, in HSA
  6. pp 467--472 (\mn@eprint {arXiv} {arXiv:1010.5680})

\bibitem[\protect\citeauthoryear{Ma{\'{\i}}z~Apell{\'a}niz
  et~al.,}{Ma{\'{\i}}z~Apell{\'a}niz et~al.}{2014}]{Maizetal14a}
Ma{\'{\i}}z~Apell{\'a}niz J.,  et~al., 2014, \mn@doi [A\&A]
  {10.1051/0004-6361/201423439}, \href
  {http://adsabs.harvard.edu/abs/2014A&A...564A..63M} {564, A63}

\bibitem[\protect\citeauthoryear{Ma{\'{\i}}z~Apell{\'a}niz
  et~al.,}{Ma{\'{\i}}z~Apell{\'a}niz et~al.}{2019}]{Maizetal19c}
Ma{\'{\i}}z~Apell{\'a}niz J.,  et~al., 2019, in HSA 10. pp 346--352 (\mn@eprint
  {arXiv} {arXiv:1810.12192})

\bibitem[\protect\citeauthoryear{Meingast, Alves  \& Lombardi}{Meingast
  et~al.}{2018}]{Meinetal18}
Meingast S.,  Alves J.,   Lombardi M.,  2018, \mn@doi [A\&A]
  {10.1051/0004-6361/201731396}, \href
  {http://adsabs.harvard.edu/abs/2018A&A...614A..65M} {614, A65}

\bibitem[\protect\citeauthoryear{Minniti et~al.,}{Minniti
  et~al.}{2010}]{Minnetal10}
Minniti D.,  et~al., 2010, \mn@doi [New Astronomy]
  {10.1016/j.newast.2009.12.002}, \href
  {https://ui.adsabs.harvard.edu/abs/2010NewA...15..433M} {15, 433}

\bibitem[\protect\citeauthoryear{Munari, Sordo, Castelli  \& Zwitter}{Munari
  et~al.}{2005}]{Munaetal05}
Munari U.,  Sordo R.,  Castelli F.,   Zwitter T.,  2005, \mn@doi [A\&A]
  {10.1051/0004-6361:20042490}, \href
  {http://adsabs.harvard.edu/abs/2005A&A...442.1127M} {442, 1127}

\bibitem[\protect\citeauthoryear{Naz{\'e}, Rauw, Czesla, Mahy  \&
  Campos}{Naz{\'e} et~al.}{2019}]{Nazeetal19}
Naz{\'e} Y.,  Rauw G.,  Czesla S.,  Mahy L.,   Campos F.,  2019, \mn@doi [A\&A]
  {10.1051/0004-6361/201935141}, \href
  {https://ui.adsabs.harvard.edu/abs/2019A&A...627A..99N} {627, A99}

\bibitem[\protect\citeauthoryear{Nishiyama, Tamura, Hatano, Kato, Tanab{\'e},
  Sugitani  \& Nagata}{Nishiyama et~al.}{2009}]{Nishetal09}
Nishiyama S.,  Tamura M.,  Hatano H.,  Kato D.,  Tanab{\'e} T.,  Sugitani K.,
  Nagata T.,  2009, \mn@doi [ApJ] {10.1088/0004-637X/696/2/1407}, \href
  {http://adsabs.harvard.edu/abs/2009ApJ...696.1407N} {696, 1407}

\bibitem[\protect\citeauthoryear{Nogueras-Lara et~al.,}{Nogueras-Lara
  et~al.}{2018a}]{Noguetal18a}
Nogueras-Lara F.,  et~al., 2018a, \mn@doi [A\&A] {10.1051/0004-6361/201732002},
  \href {http://adsabs.harvard.edu/abs/2018A&A...610A..83N} {610, A83}

\bibitem[\protect\citeauthoryear{Nogueras-Lara et~al.,}{Nogueras-Lara
  et~al.}{2018b}]{Noguetal18b}
Nogueras-Lara F.,  et~al., 2018b, \mn@doi [A\&A] {10.1051/0004-6361/201833518},
  \href {https://ui.adsabs.harvard.edu/abs/2018A&A...620A..83N} {620, A83}

\bibitem[\protect\citeauthoryear{Nogueras-Lara, Sch{\"o}del, Najarro,
  Gallego-Calvente, Gallego-Cano, Shahzamanian  \& Neumayer}{Nogueras-Lara
  et~al.}{2019}]{Noguetal19}
Nogueras-Lara F.,  Sch{\"o}del R.,  Najarro F.,  Gallego-Calvente A.~T.,
  Gallego-Cano E.,  Shahzamanian B.,   Neumayer N.,  2019, \mn@doi [A\&A]
  {10.1051/0004-6361/201936322}, \href
  {https://ui.adsabs.harvard.edu/abs/2019A&A...630L...3N} {630, L3}

\bibitem[\protect\citeauthoryear{Rieke \& Lebofsky}{Rieke \&
  Lebofsky}{1985}]{RiekLebo85}
Rieke G.~H.,  Lebofsky M.~J.,  1985, \mn@doi [ApJ] {10.1086/162827}, \href
  {http://adsabs.harvard.edu/abs/1985ApJ...288..618R} {288, 618}

\bibitem[\protect\citeauthoryear{Salas, Ma{\'{\i}}z~Apell{\'a}niz  \&
  Barb{\'a}}{Salas et~al.}{2015}]{Salaetal15}
Salas J.,  Ma{\'{\i}}z~Apell{\'a}niz J.,   Barb{\'a} R.~H.,  2015, in HSA 8. pp
  615--615 (\mn@eprint {arXiv} {arXiv:1410.6767})

\bibitem[\protect\citeauthoryear{Skrutskie et~al.,}{Skrutskie
  et~al.}{2006}]{Skruetal06}
Skrutskie M.~F.,  et~al., 2006, \mn@doi [AJ] {10.1086/498708}, \href
  {http://adsabs.harvard.edu/abs/2006AJ....131.1163S} {131, 1163}

\bibitem[\protect\citeauthoryear{Stead \& Hoare}{Stead \&
  Hoare}{2009}]{SteaHoar09}
Stead J.~J.,  Hoare M.~G.,  2009, \mn@doi [MNRAS]
  {10.1111/j.1365-2966.2009.15530.x}, \href
  {http://adsabs.harvard.edu/abs/2009MNRAS.400..731S} {400, 731}

\bibitem[\protect\citeauthoryear{Strai{\v{z}}ys \& Laugalys}{Strai{\v{z}}ys \&
  Laugalys}{2008}]{StraLaug08c}
Strai{\v{z}}ys V.,  Laugalys V.,  2008, Baltic Astronomy, \href
  {https://ui.adsabs.harvard.edu/abs/2008BaltA..17..253S} {17, 253}

\bibitem[\protect\citeauthoryear{Strai{\v{z}}ys \&
  Lazauskait{\.{e}}}{Strai{\v{z}}ys \& Lazauskait{\.{e}}}{2008}]{StraLaza08}
Strai{\v{z}}ys V.,  Lazauskait{\.{e}} R.,  2008, Baltic Astronomy, \href
  {https://ui.adsabs.harvard.edu/abs/2008BaltA..17..277S} {17, 277}

\bibitem[\protect\citeauthoryear{Suh \& Hong}{Suh \& Hong}{2017}]{SuhHong17}
Suh K.-W.,  Hong J.,  2017, \mn@doi [Journal of Korean Astronomical Society]
  {10.5303/JKAS.2017.50.4.131}, \href
  {http://adsabs.harvard.edu/abs/2017JKAS...50..131S} {50, 131}

\bibitem[\protect\citeauthoryear{Suh \& Kwon}{Suh \& Kwon}{2011}]{SuhKwon11}
Suh K.-W.,  Kwon Y.-J.,  2011, \mn@doi [MNRAS]
  {10.1111/j.1365-2966.2011.19462.x}, \href
  {https://ui.adsabs.harvard.edu/abs/2011MNRAS.417.3047S} {417, 3047}

\bibitem[\protect\citeauthoryear{Wang \& Chen}{Wang \& Chen}{2019}]{WangChen19}
Wang S.,  Chen X.,  2019, \mn@doi [ApJ] {10.3847/1538-4357/ab1c61}, \href
  {https://ui.adsabs.harvard.edu/abs/2019ApJ...877..116W} {877, 116}

\makeatother
\end{thebibliography}

%
%
%
%



\appendix
\section{Non-linear photometric effects caused by extinction in the NIR}

\begin{figure}
\centerline{\includegraphics[width=\linewidth]{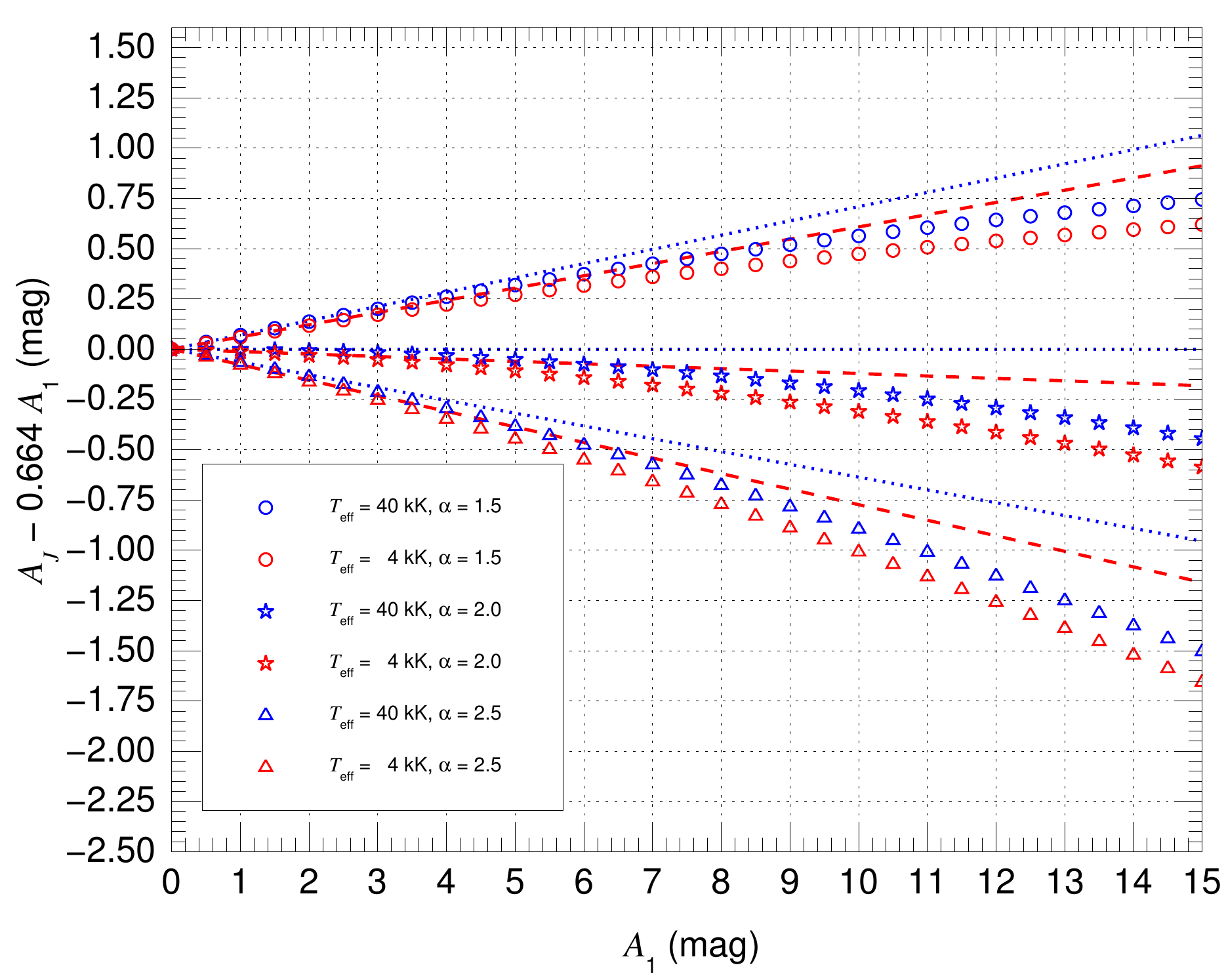}}
\centerline{\includegraphics[width=\linewidth]{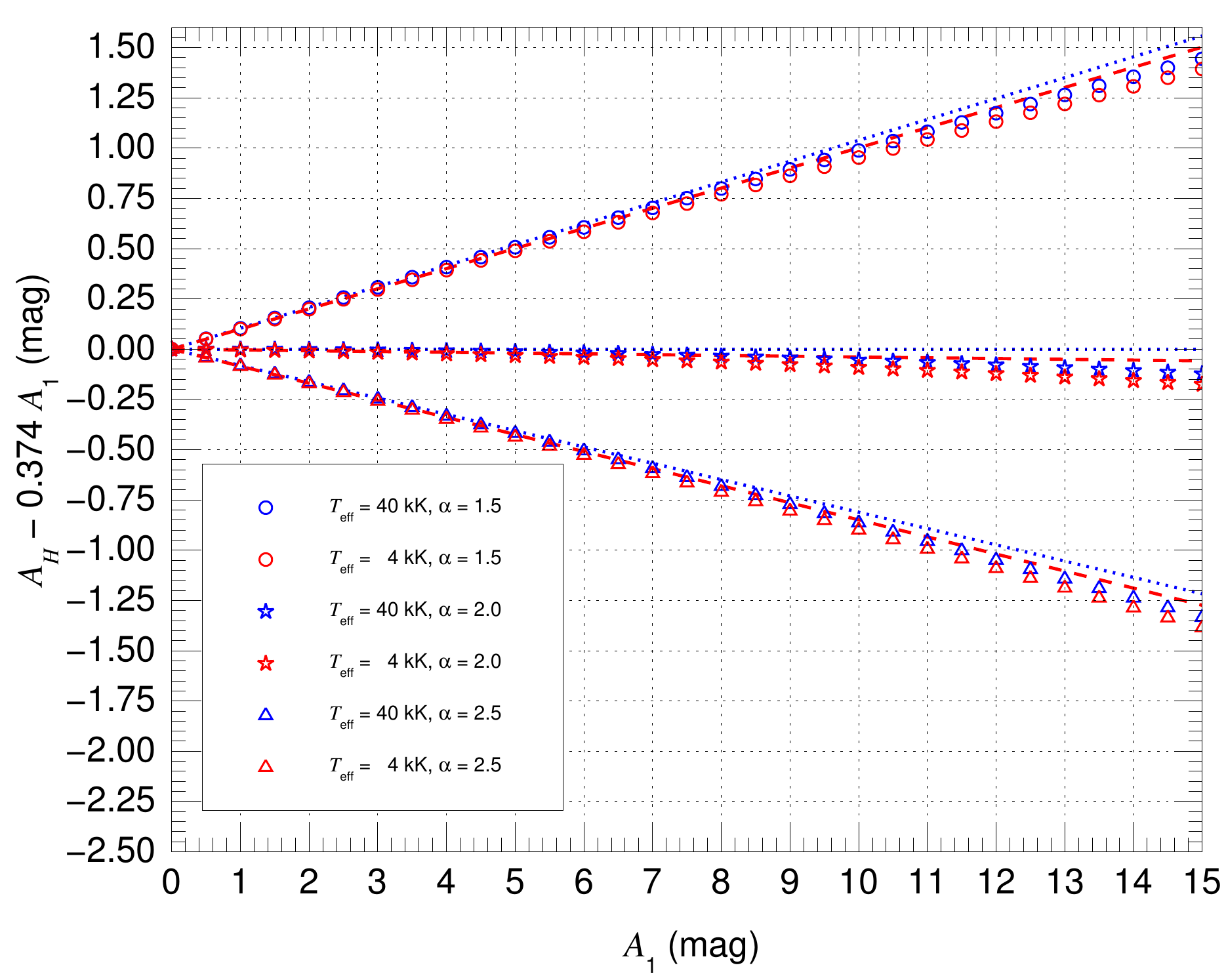}}
\centerline{\includegraphics[width=\linewidth]{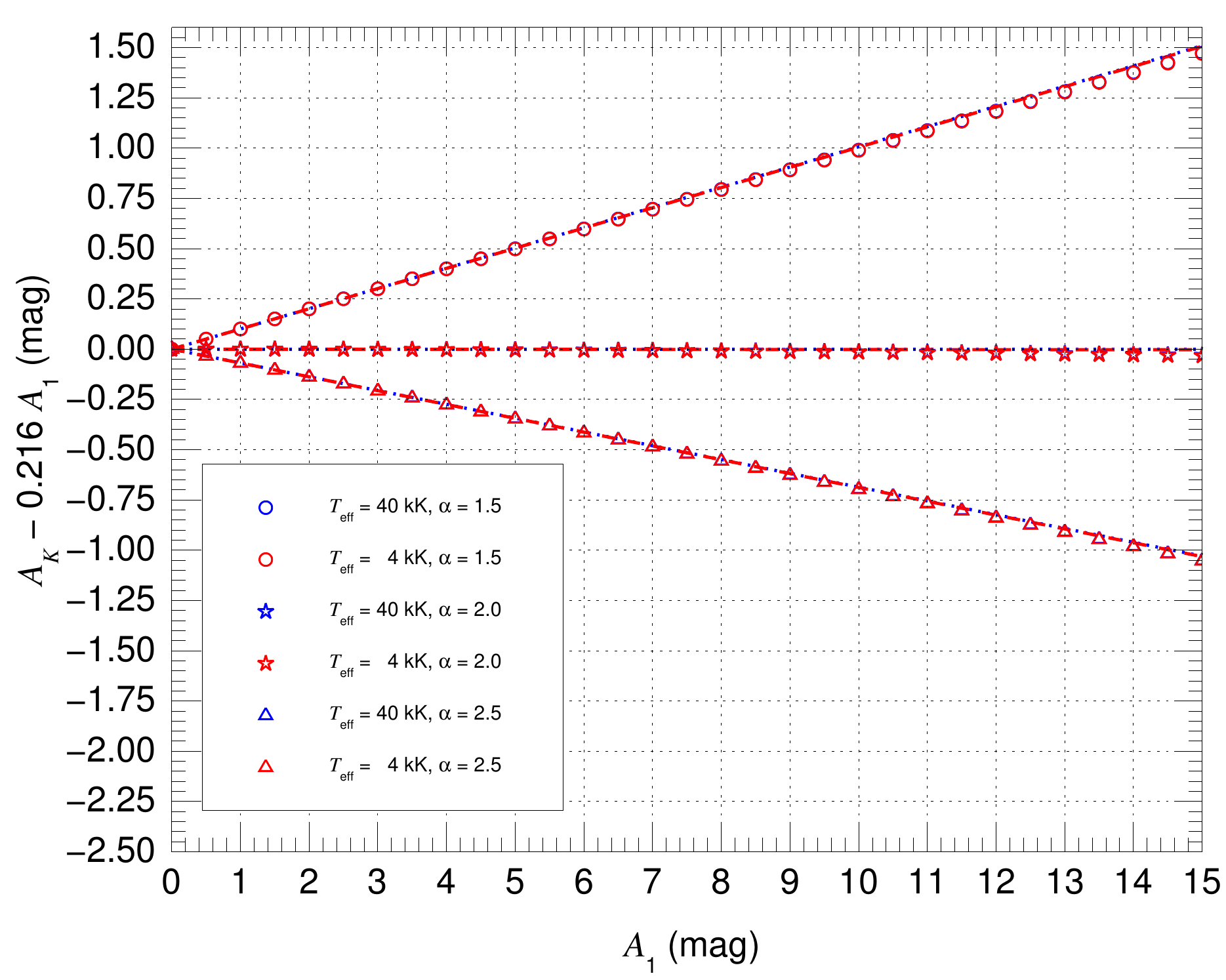}}
 \caption{Band-integrated extinctions ($A_J$, $A_H$, and $A_K$) as a function of the monochromatic amount of extinction $A_1$ for the combination of 
         two effective \Teff\ (4~kK and 40~kK) and three values of $\alpha$ (1.5, 2.0, and 2.5). Symbols are used for the actual values and lines
         for the extrapolation from the low-extinction regime. The coefficient multiplying $A_1$ used to modify the ordinate has been chosen to make the
         behavior for the \Teff~=~40~kK, $\alpha$~=~2.0 case flat for the low-extinction regime.}
\label{extinction}
\end{figure}

\begin{figure}
\centerline{\includegraphics[width=\linewidth]{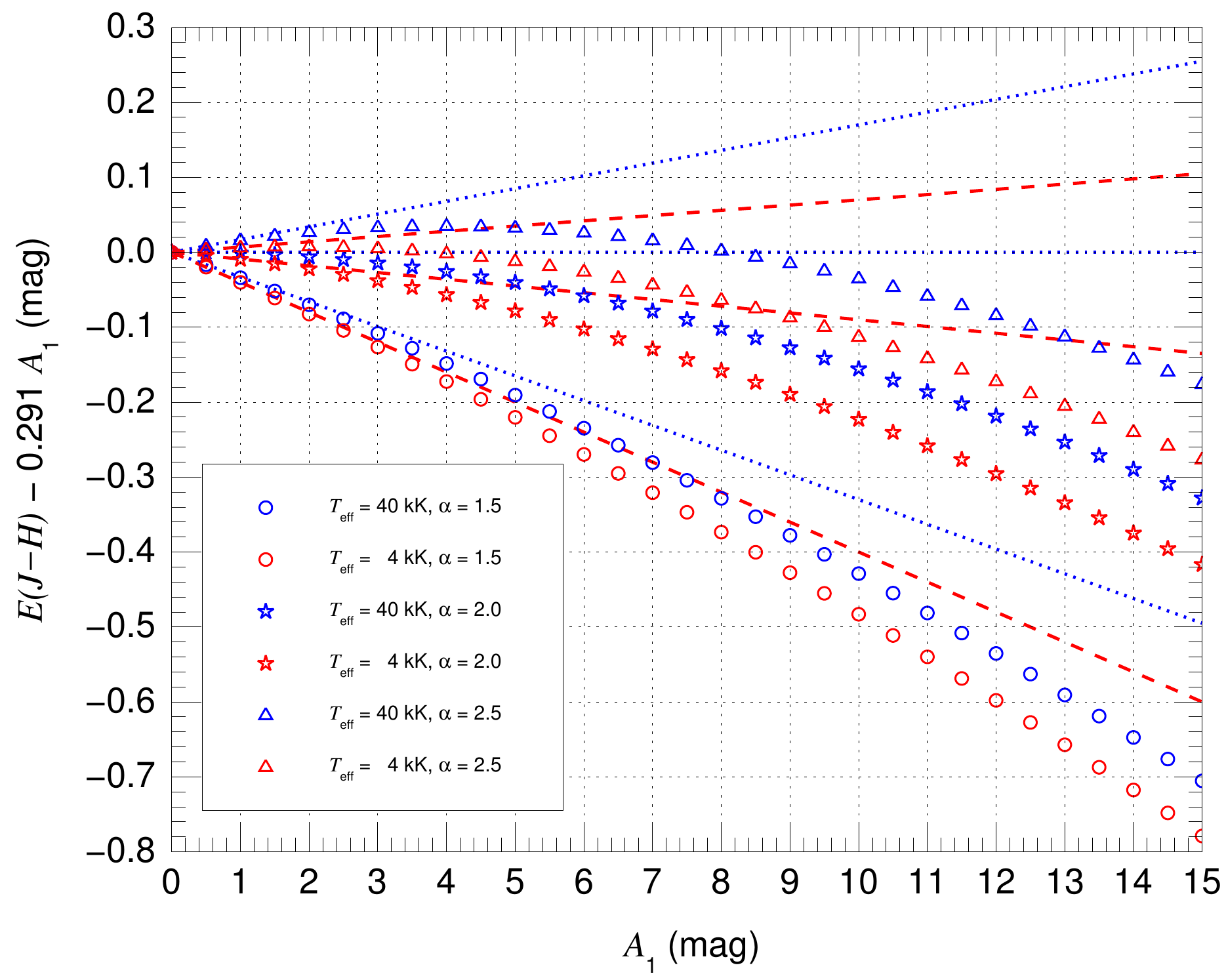}}
\centerline{\includegraphics[width=\linewidth]{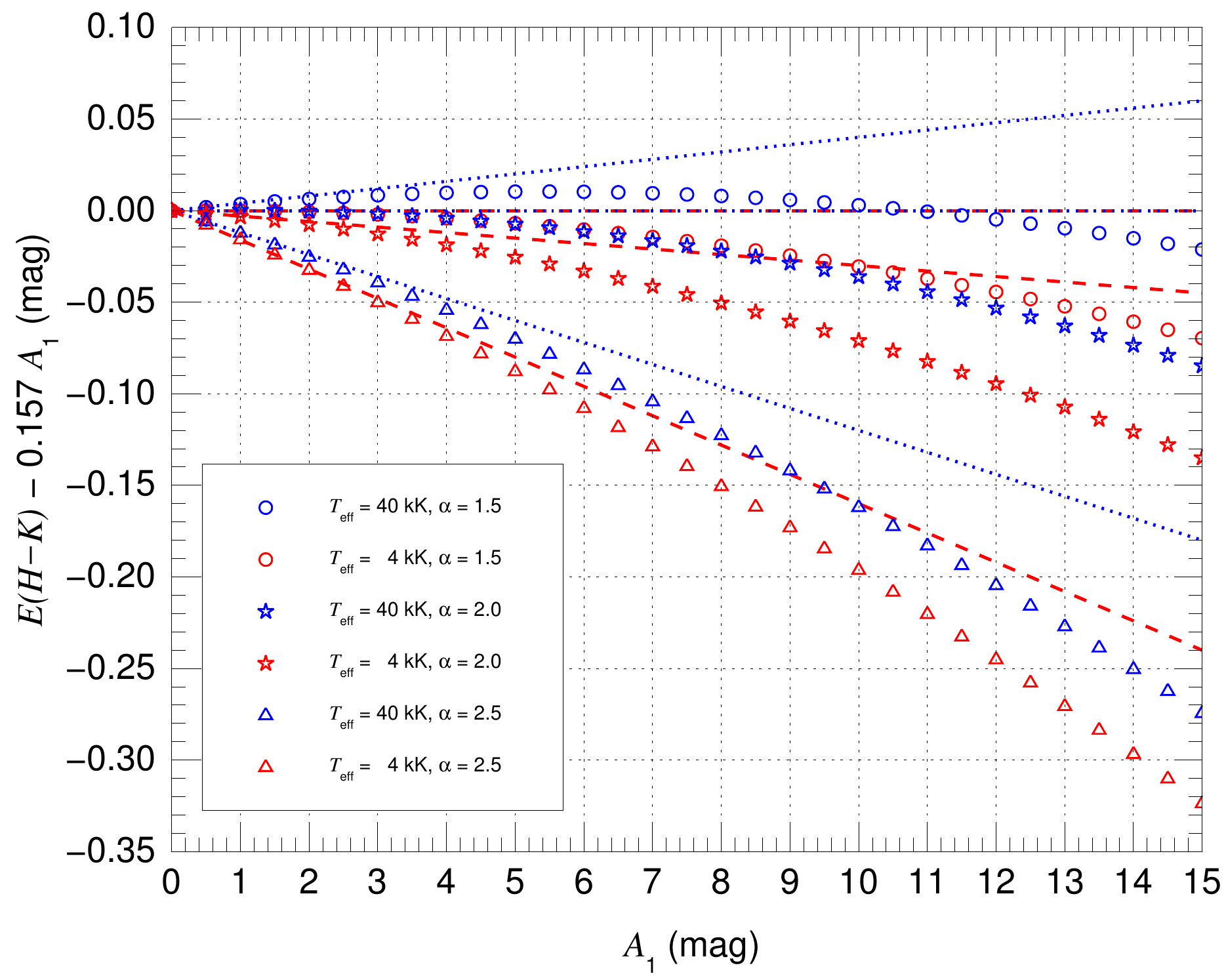}}
\caption{Band-integrated colour excesses [$E(J-H)$ and $E(H-K)$] as a function of the monochromatic amount of extinction $A_1$ for the combination of 
         two effective \Teff\ (4~kK and 40~kK) and three values of $\alpha$ (1.5, 2.0, and 2.5). Symbols are used for the actual values and lines
         for the extrapolation from the low-extinction regime. The coefficient multiplying $A_1$ used to modify the ordinate has been chosen to make the
         behavior for the \Teff~=~40~kK, $\alpha$~=~2.0 case flat for the low-extinction regime. Note that for $E(H-K)$ the 
         \Teff~=~40~kK, $\alpha$~=~2.0 and \Teff~=~4~kK, $\alpha$~=~1.5 cases are nearly coincident for the low-extinction regime.}
\label{colorexcess1}
\end{figure}

\begin{figure}
\centerline{\includegraphics[width=\linewidth]{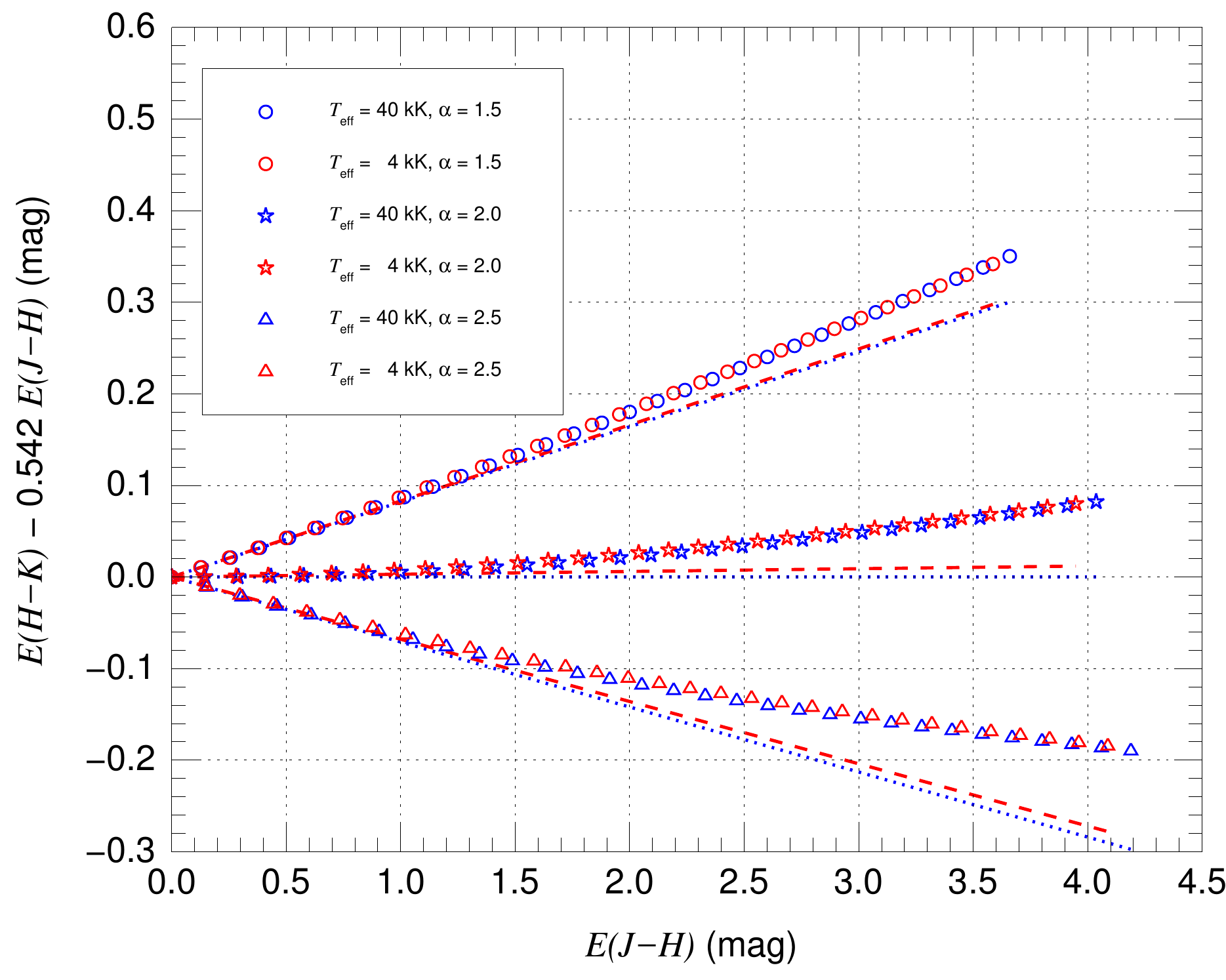}}
\caption{Band-integrated colour excess $E(H-K)$ as a function of $E(J-H)$ for the combination of two \Teff\ (4~kK and 40~kK) and three values 
         of $\alpha$ (1.5, 2.0, and 2.5). Symbols are used for the actual values and lines for the extrapolation from the low-extinction regime. The 
         coefficient multiplying $E(J-H)$ used to modify the ordinate has been chosen to make the behavior for the \Teff~=~40~kK, $\alpha$~=~2.0 case 
         flat for the low-extinction regime. Symbols are plotted for the same values of $A_1$ as in Figs.~\ref{extinction}~and~\ref{colorexcess1} i.e. a range
         of 0~mag to 15~mag at 0.5~mag steps.}
\label{colorexcess2}
\end{figure}

$\,\!$\indent The non-linear behavior of photometry with extinction (also known as bandwidth effects) in the optical has been known as far back as
\citet{Blan56,Blan57} and in previous papers we have provided a modern analysis of the associated issues \citep{Maiz04c,Maiz13b,Maizetal14a,MaizBarb18},
to which the reader is referred for notation and basic concepts. In the NIR, bandwidth effects were first studied by \citet{JoneHyla80} and 2MASS analyses 
were given by \citet{StraLaza08} and by \citet{SteaHoar09}. Here we provide another 2MASS analysis adapted to the circumstances on this paper and with several plots 
(Figs.~\ref{extinction},~\ref{colorexcess1},~and~\ref{colorexcess2}) to illustrate the non-linearity of extinction and colour excesses.

The non-linearity of photometry with extinction has three main effects:

\begin{itemize}
 \item {\bf Monochromatic parameters are a must.} The amount and type of extinction in an extinction law have to be defined through monochromatic quantities 
       i.e. quantities that do not depend on an integral over wavelength. Every time you see an extinction law whose type (e.g. dust grain size) is defined by 
       $R_V$ or its amount of extinction by $A_V$ or $E(B-V)$ you should raise your hand, protest, and ask for a monochromatic equivalent such as \RV, 
       $A_{5495}$, or \EBV. Otherwise, you are not measuring the type or amount of extinction directly but a combination of those with the effect of the 
       input SED. In this paper the type of extinction is defined by $\alpha$, a direct parameter of the extinction law, and 
       the amount of extinction by $A_1$, the extinction at a single wavelength. Keep in mind that central wavelengths, sometimes used to define filters, are 
       also band-integrated.
 \item {\bf Non-linearity with input SEDs.} If you take two stars, one (intrinsically) red and one (intrinsically) blue, and you place the same amount and 
       type of dust in front of them, you are going to get different band-integrated extinctions $A_X$, colour excesses $E(X-Y) \equiv A_X - A_Y$, and other 
       band-integrated parameters which are a combination of those (e.g. $R_V = A_V/E(B-V)$). 
 \item {\bf Non-linearity with the amount of extinction.} Take one star and place a given amount of dust in front of it to get an extinction $A_X$ and a colour
       excess $E(X-Y)$. Now, double the amount of dust (maintaining the type) and what do you get? Not exactly $2\,A_X$ and $2\,E(X-Y)$, as for the added 
       extinction you have an input SED which is effectively redder. 
\end{itemize}

With these points in mind, we test the behavior of band-integrated quantities composed of the three 2MASS passbands $J$, $H$, and $K$ using as input SEDs a
blue and a red star and varying the type of extinction with $\alpha$ (1.5,~2.0,~and~2.5) and the amount with $A_1$ (between 0 and 15). For the blue and red
stars we respectively use a 40~kK and a 4~kK giant SEDs with solar metallicity from the \citet{Maiz13a} grid, as representative of the two extremes of the
population of high-luminosity extinguished objects in the Galaxy. More specifically, the blue SED uses a TLUSTY \citep{LanzHube03,LanzHube07} model in the 
optical (which is not relevant here) and a \citet{Munaetal05} in the NIR. The reason for using the Munari models in the range of interest here is that the 
TLUSTY models do not yield the correct NIR colours for zero extinction when comparing with real data but the \citet{Munaetal05} ones do. The red SED is from the 
MARCS models \citep{Gustetal03}, as are the red giant SEDs used elsewhere in this paper. We have selected a giant model for a better 
comparison with the objects in this paper but note that the dominant effect is \Teff, not gravity\footnote{More specifically, we use the term 
``giant'' to refer to a luminosity class of 3.0, i.e. the red crosses in Fig.~2 of \citet{Maiz13a}. In that respect, a 4.75~kK giant is at the red clump
location while a 4~kK giant is halfway up the red giant branch, so the second has a significantly higher luminosity and lower gravity.}. This is especially so 
for the 40~kK model, as O-star SEDs are mostly invariant with respect to gravity if one excludes wind effects (which are not present in the models used here). 

The SEDs have been extinguished first and then
their synthetic photometry calculated with the code created for CHORIZOS \citep{Maiz04c} using the 2MASS zero points of \citet{MaizPant18}. Keep in mind
that some SED-fitting codes do not do this correctly, as they only extinguish fluxes at central wavelengths and in that way do not take into account 
non-linear photometric effects.

In Figure~\ref{extinction} we plot the band-integrated $A_J$,~$A_H$,~and~$A_K$ extinctions as a function of the amount of extinction. In the vertical axes we
have subtracted a linear component (described in the caption) to flatten the plots for a better comparison. Starting with the bottom plot, we see that the
symbols essentially (a) coincide with the linear extrapolation from the low-extinction regime shown by the lines and (b) do not show differences between
the blue and red SEDs. In other words, the behavior of the extinction in the $K$ band is well approximated by a linear model at least up to $A_1$~=~15~mag.
The different trends in $A_K$ as a function of $A_1$ for different values of $\alpha$ are just a direct manifestation of the different values of $A_K/A_1$ 
as a function of $\alpha$. For example, that value is 0.216 for $\alpha = 2.0$, higher for $\alpha = 1.5$, and lower for $\alpha = 2.5$.

Continuing with the middle plot of Fig.~\ref{extinction} we see two differences between $A_H$ and $A_K$. First, the blue and red points show a small offset, 
indicating there is a non-linearity with input SED (second effect above). Second, the points (measured values) and the lines (extrapolations from the
low-extinction regime diverge, indicating there is a non-linearity with the amount of extinction (third effect above). These non-linear effects are small for
$A_H$ in this range for $A_1$, so we can call this a weak non-linear regime. However, they become much larger for $A_J$ (top plot), where we enter the 
strong non-linear regime. This progression from difficult to see ($K$) to strong ($J$) through weak ($H$) is attributed to the different amounts of extinction
in the three bands represented by the three linear components subtracted in the vertical axes (which are the $A_X/A_1$ values for the blue SED in the
low-extinction regime): 0.216 for $K$, 0.374 for $H$, and 0.664 for $J$. Note that eventually all equivalent plots become non-linear. What happens with $K$ is
that in this $A_1$ range (selected from the properties of our sample i.e. 2MASS stars with good-quality photometry) there is not enough extinction for
significant non-linear effects to appear.

As an example of how non-linear effects are ignored in some of the previous papers on extinction, let us examine the often referenced extinction law of 
\citet{Nishetal09}. Those authors determine that the extinction law towards the Galactic Center has $\alpha = 2.0$ and that the following 
ratios\footnote{They use the IRSF definitions instead of the 2MASS ones but that is relevant only to the specific values, not to their variation.} apply: 
$A_J/A_K = 3.02\pm 0.04$ and $A_H/A_K = 1.73\pm 0.03$. When one applies the values in Fig.~\ref{extinction} for a 4~kK giant with $\alpha = 2.0$ the first 
ratio goes from 3.02 at low extinction values to 2.92 for $A_1$~=~15~mag (a 3\% effect) while the second ratio goes from 1.713 at low extinction values to 1.693 
for $A_1$~=~15~mag (a significantly smaller 1\% effect in agreement with the previous paragraph). If we repeat the exercise for a 40~kK giant, $A_J/A_K$ goes
from 3.08 to 2.97 and $A_H/A_K$ from 1.729 to 1.707. The \citet{Nishetal09} values are in the correct ballpark but those extinction ratios are a function of 
\Teff\ and $A_1$, not constant, for a given extinction law.

We continue with Fig.~\ref{colorexcess1}, which is essentially a subtraction of the first and second panels of Fig.~\ref{extinction} (top panel) and of
the second and third panels of the same Figure (bottom panels). This can be seen in the coefficients that multiply $A_1$ in the vertical axes (allowing
for round-off in the last digit). As a consequence, we see that non-linear effects in $J-H$ are about twice as strong as in $H-K$ (note the different scales in
the vertical axes), as expected by the previous analysis. 
{Therefore, {\bf if we assume an extinction law, the same value of $E(J-H)$ or of $E(H-K)$ does not correspond to the same $A_1$}, 
as the precise value for $A_1$ for a given color excess also depends on \Teff. Also, {\bf doubling $E(J-H)$ or of $E(H-K)$ does not exactly
double $A_1$} or any other monochromatic measurement of extinction.} 

We finish with Fig.~\ref{colorexcess2}, which is a modified version of the classic $H-K$ vs. $J-H$ colour-colour diagram with two important differences. First, we 
have subtracted the intrinsic colours, implying that we know the input SED for each star, as $E(J-H) = (J-H) - (J-H)_0$ and $E(H-K) = (H-K) - (H-K)_0$. Second,
the vertical axis is a linear combination of $E(H-K)$ and $E(J-H)$ as opposed to simply $E(H-K)$. This is done to flatten the vertical axis for the case of
the 40~kK star with $\alpha = 2.0$, as we have done for previous plots. The importance of this plot is that is built from colour excesses alone, with no
knowledge of $A_1$, albeit requiring a knowledge of the input SED. In that way, it can be built from observational information alone without modelling
extinction.

In Fig.~\ref{colorexcess2} we see different trends as a function of $\alpha$, something that also appears in the previous two figures and in the data in 
Fig.~\ref{2MASS_cc_05}, which we exploit in this paper to simultaneously measure $A_1$ and $\alpha$. There is, however, a difference regarding
non-linearity. While the trajectories for blue and red stars are clearly curved and deviate from the low-extinction extrapolation (third effect above), they do
not show large differences as a function of \Teff. More specifically, the trajectories with the same $\alpha$ for different temperatures show a scatter
of less than 0.01~mag in the vertical axis but show small deviations in the location of a given value of $A_1$ in the horizontal axis (as we already know that
$E(J-H)$ is not a direct measurement of $A_1$). This is important because it means that the extinction trajectories in a $H-K$ vs. $J-H$ colour-colour diagram
will be slightly curved but ``quasi-parallel'' (more appropriately, similar but displaced with respect to their origins) for different \Teff.
Therefore, the extinction trajectories of blue and red stars will tend not to cross, something that would reduce the value of colour-colour diagrams to discriminate
between objects of different temperatures. As a counter-example of this effect in the optical, where non-linearity effects due to extinction are much stronger,
the reader is referred to the left panel of Fig.~1 in \citet{Maiz04c}. There, a star with \Teff~=~10~kK and \EBV~=~1.0~mag has nearly identical 
$U-B$ and $B-V$ colours as a \Teff~=~5~kK with no extinction (i.e. the extinction trajectory of the hotter star has taken it to overlap with the cooler
star). However, as we increase extinction for both, their trajectories diverge. Why does that happen if at that point they have the same colours? Because that
is not all there is to it, as "same colours" does not mean "same SEDs" and the detailed wavelength dependence of each object is different. The results here
indicate that this effect is weak or non-existent in the $H-K$ vs. $J-H$ plane.

As an exercise for the reader, we leave the option of repeating the analysis starting in Fig.~\ref{2MASS_cc_05} but replacing the vertical axis by 
$(H-K)-0.542\,(J-H)$. Doing so flattens the vertical axis and leads to an observational plot where the effects of amount and type of extinction approximately
correspond to the $x$ and $y$ axes, respectively. The word ``approximately'' should be emphasized here, given the effect of the different input SEDs and of
photometric non-linearity.


\bsp	
\label{lastpage}
\end{document}